\let\oldvec\vec% Store \vec in \oldvec
\documentclass[smallextended]{svjour3}       % onecolumn (second format)
\let\vec\oldvec% Restore \vec from \oldvec

\smartqed  % flush right qed marks, e.g. at end of proof

\usepackage[T1]{fontenc}
\usepackage[utf8]{inputenc}
\usepackage{lmodern}
\usepackage[english]{babel}

\usepackage{graphicx}
\usepackage{amssymb}
\usepackage{color}
\usepackage[]{natbib}
\usepackage{amsmath}
\usepackage{graphicx}
\usepackage{epsfig}
\usepackage{amsfonts}
\usepackage[titletoc,title]{appendix}
\usepackage{seqsplit}
\usepackage{lipsum}
\usepackage{enumerate}
\usepackage[footnotesize]{caption}
\usepackage{hyperref}

\graphicspath{{./images/}}
\usepackage{longtable}
\usepackage[left=1.0in, right=1.0in, top=1.0in, bottom=1.0in, includehead,includefoot]{geometry}

\numberwithin{equation}{section}

\usepackage{eurosym}
\usepackage{mwe}    % loads »blindtext« and »graphicx«
\usepackage[subrefformat=parens,labelformat=parens]{subfig}
\usepackage{chemformula}
\let\ce\ch
\usepackage{verbatim}
\usepackage{changepage}
\usepackage{array}
\usepackage{xcolor}

\newcolumntype{+}{!{\vrule width 2pt}}

\newlength\savedwidth

\newcommand\thickhline{\noalign{\global\savedwidth\arrayrulewidth\global\arrayrulewidth 2pt}%
\hline
\noalign{\global\arrayrulewidth\savedwidth}}

\newcommand{\ca}{\ce{Ca^2+}\xspace}
\newcommand{\mms}{\mu \mathrm{m}^2 \mathrm{s}^{-1}\xspace}
\newcommand{\cms}{\mathrm{cm}^2 \mathrm{s}^{-1}\xspace}
\newcommand{\s}{\mathrm{s}\xspace}
\newcommand{\is}{\mathrm{s}^{-1}\xspace}
\newcommand{\mM}{\mu \mathrm{M}\xspace}
\newcommand{\mm}{\mu \mathrm{m}\xspace}
\newcommand{\ip}{\ce{InsP_3}\xspace}
\newcommand{\iprs}{\ce{InsP_3Rs}\xspace}
\newcommand{\ipr}{\ce{InsP_3R}\xspace}

\newcommand{\PD}[2]{\frac{\partial #1}{\partial #2}}

\definecolor{acolor}{rgb}{0.30,0.03,0.66}

\usepackage{lineno}

\journalname{}

\begin{document}
\title{Adhesion-driven patterns in a calcium-dependent model of cancer cell movement}

\titlerunning{Mathematical model of cancer and calcium}        % if too long for running head

\author{Kaouri, K \and
        Bitsouni, V         \and
        Buttenschön, A \and
        Thul, R
}

\institute{ K.\ Kaouri \at
              School of Mathematics, Cardiff University, CF24 4AG, UK\\
              Tel.: +442920875259\\
              \email{KaouriK@cardiff.ac.uk}
         \and
         V.\ Bitsouni \at
              SciCo Cyprus, Nicosia 1700, Cyprus \& School of Mathematics, Cardiff University, CF24 4AG, UK \\
              \email{vbitsouni@gmail.com}
           \and
           A.\ Buttenschön \at
           Department of Mathematics, University of British Columbia, Vancouver V6T 1Z2, BC, Canada\\
         \email{abuttens@math.ubc.ca}
           \and
          R.\ Thul \at
           School of Mathematical Sciences \& Centre for Mathematical Medicine and Biology \\
           University of Nottingham, Nottingham, NG7 2RD, UK\\ \email{ruediger.thul@nottingham.ac.uk}
      }

\date{Received: date / Accepted: date}

\maketitle

\begin{abstract}
Cancer cells exhibit increased motility and proliferation, which are instrumental in the formation of tumours and metastases. These pathological changes can be traced back to malfunctions of cellular signalling pathways, and calcium signalling plays a prominent role in these.  We formulate a new model for cancer cell movement which for the first time explicitly accounts for the dependence of cell proliferation and cell-cell adhesion on calcium. At the heart of our work is a non-linear, integro-differential (non-local) equation for cancer cell movement, accounting for cell diffusion, advection and proliferation. We also employ an established model of cellular calcium signalling with a rich dynamical repertoire that includes experimentally observed periodic wave trains and solitary pulses. The cancer cell density exhibits travelling fronts and complex spatial patterns  arising from an adhesion-driven instability (ADI). We show how the different calcium signals and variations in the strengths of cell-cell attraction and repulsion shape the emergent cellular aggregation patterns, which are a key component of the metastatic process. Performing a linear stability analysis, we identify parameter regions corresponding to ADI. These regions are confirmed by numerical simulations, which also reveal different types of aggregation patterns and these patterns are significantly affected by \ca. Our study demonstrates that the maximal cell density decreases with calcium concentration, while the frequencies of the calcium oscillations and the cell density oscillations are approximately equal in many cases. Furthermore, as the calcium levels increase the speed of the travelling fronts increases, which is related to a higher cancer invasion potential. These novel insights provide a step forward in the design of new cancer treatments that may rely on controlling the dynamics of cellular calcium.

\end{abstract}

\section{Introduction}
\label{intro}
%KK: I think the following part is not needed, I am commenting out
%The cell cycle is a highly controlled multi-stage process, requiring mitogenic signals, which are transmitted into the cell by the transmembrane receptors and bind signalling molecules, such as cell-cell adhesion molecules \citep{hanahan2000hallmarks}.
Cell-cell adhesion and cellular proliferation are fundamental features of multicellular organisms, along with cell division, migration and apoptosis. These processes are orchestrated and coordinated by a multitude of cellular signalling pathways \citep{alberts2000molecular}. When these signalling cascades are disturbed, numerous pathologies ensue, including cancer. Amongst the many molecular changes that characterise cancer, alterations of intracellular calcium (\ca) signalling have been identified as a crucial driver \citep{colomer2007physiological}. In particular, \ca  has been reported as a key factor in cellular proliferation \citep{roderick20082,shapovalov2013calcium} and in cellular adhesion \citep{weinberg2013biology}. Here, we formulate and analyse for the first time a model that describes the evolution of a cancer cell density incorporating the effects of \ca in the adhesion and proliferation processes.

%KK: I think the first part of the paragraph can be omitted, I am commenting it out
%In cancer cells there is often significant remodelling in the expression of \ce{Ca^2+}-involved proteins, as well as a sensitivity of oncogenic mechanisms to regulation by specific  \ce{Ca^2+}-ions \citep{monteith2012calcium,prevarskaya2014remodelling}.
Rising levels of intracellular \ca  have been shown to increase the proliferation of cancer cells in various cancer types such as breast and prostate cancer, melanoma, hepatocellular and non-small-cell lung carcinoma \citep{prevarskaya2014remodelling,prevarskaya2018ion}.  Experiments \citep{simpson1986calcium,taylor1992inhibition} have shown that increasing extracellular \ca levels increased intracellular calcium \ca levels, which increased the cell number and the DNA synthetic ability of cell lines.

Cellular adhesion is mediated through cadherins, which are transmembrane proteins and  belong to the class of calcium-dependent cell adhesion molecules (CAMs) \citep{weinberg2013biology}. As an example, consider epithelial cells, which bind to each other by linking the extracellular domains of E-cadherins \citep{morales2002hallmarks}. The cytosolic domain of E-cadherin binds to $\beta-$catenin, which in turn binds to the cytoskeleton. Changes in the function of $\beta-$catenin result in the loss of the ability of E-cadherin to sustain sufficient cell–cell adhesion \citep{makena2020subtype,wijnhoven2000cadherin}, while alterations in any type of cadherin expression may affect cell adhesion and signal transduction \citep{cavallaro2002cadherins}. Intracellular \ca directly impacts on the dynamics of both cadherins and catenins \citep{ko2001cell}.  Moreover, \cite{hills2012calcium} have shown that activation of extracellular \ca-sensing receptors leads to an increase in E-cadherin expression and an increase in the binding of $\beta-$catenin. In cancer, disrupted cell-cell adhesion due to abnormal expression of cadherins and their associated catenins has been linked to metastasis \citep{morales2002hallmarks}. For instance, \citep{byers1995role,cavallaro2004cell} have shown a reduced expression of cadherins in various cancer types, including melanoma, prostate, breast cancer, invasive carcinomas and carcinoma cell lines, and cancers of epithelial origin, when \ca levels are increased. This results in  a reduced force between cells and consequently to cell migration. These results are in line with findings that show that altering CAM function in metastatic cancer cells blocked their ability to invade healthy tissue and move to secondary sites \citep{kotteas2014intercellular,naik2008attenuation,slack2009vascular,zhu1992blocking}. Taken together, the combined changes in cell-cell adhesion and the increase in the proliferation rate and their dependence on \ca are important mechanisms in cancer and enhance the formation of cancer cell clusters/aggregations that can migrate in a collective manner, a process critical for cancer progression \citep{friedl2004collective,glinsky2003intravascular,knutsdottir2014mathematical}.

\ce{Ca^2+} signalling uses an extensive molecular repertoire of signalling components termed the  \ce{Ca^2+} signalling ``toolkit" \citep{Berridge:2000jn}. A key feature of \ca signalling is \ca release from the Endoplasmic Reticulum (ER) to the cytosol through inositol-1,4,5-trisphosphate (\ip) receptors (\iprs). Together with \ca resequestration from the cytosol through sarco-endoplasmic \ca ATPase (SERCA) pumps, a process known as calcium-induced-calcium release can give rise to intracellular \ca oscillations \citep{Berridge:1988vz,Berridge:2000jn,Parekh:2011dq,Dupont:2016js,Thul:2008uu,Dupont:2011fj,Dupont:2016fc,Schuster:2002uh,Uhlen:2010cs,Powell:2020ji,Sneyd:2017fs}. In addition, \ca can spread across a population of cells, forming an intercellular \ca wave (\cite{bereiter2005mechanics, charles1993mechanisms, charles1991intercellular, charles1992intercellular, deguchi2000spatiotemporal, narciso2017release, sanderson1981ciliary, yang2009effects, young1999calcium}).

%The biological process whereby calcium is able to activate calcium release from intracellular \ce{Ca^2+} stores is known as Calcium Induced Calcium Release (CICR) and it is  a widely occurring cellular signalling process, crucial for excitation-contraction coupling in cardiac muscle and present in many other non-muscle cells \citep{dupont2016models}.

Mathematical models of intracellular \ca oscillations vary substantially in their complexity, ranging from two coupled nonlinear ordinary differential equations (ODEs) to three-dimensional hybrid partial differential equations (PDEs) --- see \citep{Dupont:2016fc,Falcke:2018jr} for recent perspectives. In the present study, we employ the model developed in \citep{atri1993single}, which for simplicity we will call the `Atri model'. The Atri model is a so-called `minimal' model consisting of only two ODEs that can generate non-linear relaxation oscillations at constant \ip concentrations \citep{dupont2016models,keener2009mathematical1,keener2009mathematical2}. Importantly, the Atri model most consistently described hormone-induced \ca oscillations in HeLa cells (an immortal cell line derived from cervical cancer cells), compared to seven other minimal models for intracellular \ca oscillations \citep{estrada2016cellular}. In addition, the mathematical structure of the Atri model allows us to determine analytically the parameter range sustaining calcium oscillations and other bifurcations of the system --- see \citep{atri1993single, kaouri2019simple}. Despite its  simplicity, the Atri model generates prototypical \ce{Ca^2+} signals such as \ca oscillations and action potentials which correspond to periodic wave trains  and solitary pulses, respectively, when \ca diffusion is taken into account. The Atri model is, hence, sufficient for our modelling framework since our focus is on studying cancer cell movement with \ca signals as input.

We base our model for the cancer cell density on previously published work (\cite{armstrong2006continuum,bitsouni2017mathematical,bitsouni2018aggregation,bitsouni2018non,chaplain2011mathematical,JMB2015,domschke2014mathematical,eftimie2017pattern,gerisch2008mathematical,GerischPainter2010_CellAdhesion,green2010non,hillen2019nonlocal,painter2015nonlocal,shuttleworth2019multiscale,szymanska2009mathematical}).  These models include nonlinear PDEs with reaction terms for cell growth/proliferation and a non-local advection term, describing cell-cell adhesion. The latter is expressed as an integral term that describes how a cell at position $\mathbf{x}$ adheres to other cells at position $\mathbf{x\pm s}$, for some $\mathbf{s}>0$ within the cell's sensing radius \citep{armstrong2006continuum}. In the present work, both the rate of cell proliferation and the strength of adhesion are taken to be \ca-dependent. It is worth noting that additional molecular components and processes could be included. For instance, integrins and TGF-$\beta$ proteins are explicitly represented in \citep{bitsouni2018aggregation,engwer2017structured} and \citep{bitsouni2017mathematical,eftimie2017pattern}, respectively. Moreover, collagen-controlled cell-matrix adhesion, where \ca is considered as constant, has been developed  in \citep{shuttleworth2019multiscale}, while \citep{ramis2008modeling,ramis2009multi} studied cadherin-dependent cellular adhesion in an individual-cell-based multiscale model. However, since our study explores the impact of intracellular \ca on cancer cell movement, we focus on diffusion, cell-cell adhesion and proliferation, the core components of cancer cell behaviour.

The structure of the paper is as follows. In Section~\ref{model_derivation} we formulate a new model that captures the crucial role of \ce{Ca^2+} signalling in cancer by incorporating \ce{Ca^2+}-dependent adhesion and proliferation effects. In Section~\ref{analytical_results}  we perform a linear stability analysis and show the ability of the model to generate ADIs and hence cell aggregations. In Section~\ref{numerics} we solve the model numerically. We present various types of aggregation patterns, as well as travelling wave patterns. Taken together, our work provides new insights into the connection between \ca  signalling and cancer cell movement, and suggests a mechanistic approach that can contribute to developing \ca-transport-targeting tools for cancer diagnosis and treatment \citep{prevarskaya2013targeting,prevarskaya2014remodelling}.
%%%%%%%%%%%%%%%%%%%%%%%%%%%%%%%%%%%%%%%%%%%%%%%%%%%%%%%%%

\section{A non-local model for calcium signalling in cancer}
\label{model_derivation}
%We have developed a novel model that describes the behaviour of a cancer cell population by taking into account the effect of \ca dynamics on cancer cell proliferation and adhesion. The cancer cell equation follows the structure of several previous models \citep{bitsouni2017mathematical,bitsouni2018aggregation}. Calcium dynamics are described by the well-established Atri model, which has been developed to describe the intracellular calcium oscillations in Xenopus oocytes and subsequently for different cell types, including glial cells \citep{wilkins1998intercellular}, mammalian spermatozoa \citep{olson2010model}, and keratinocytes \citep{kobayashi2014mathematical,kobayashi2016mathematical}. %For $T>0$ let $\Omega_T=\Omega\times\left(0, T\right)$, where $\Omega\subset\mathbb{R}$.  -\textbf{KK: since \Omega defined later when the sensing radius is explained consider moving this sentence later.}

We denote by $u\left(x,t\right)$ the cancer cell density, by $c\left(x,t\right)$ the cytosolic \ca concentration and $h\left(x,t\right)$ is the fraction of \iprs on the ER that have not been inactivated by \ca. Then the model takes the form
\begin{subequations}
\label{nonlocal_model}
\begin{align}
\dfrac{\partial c}{\partial t}&=D_c\dfrac{\partial^2 c}{\partial x^2}+J_\mathrm{ER}-J_\mathrm{pump}\,,\label{nonlocal_model;a}\\
\tau_h\dfrac{\partial h}{\partial t}&=\dfrac{k_2^2}{k_2^2+c^2}-h\,,\label{nonlocal_model;b}\\
\dfrac{\partial u}{\partial t}&=D_u\dfrac{\partial^2 u}{\partial x^2}-\dfrac{\partial}{\partial x}\left(uF\left[c,u\right]\right)+f\left(c,u\right),\label{nonlocal_model;c}
\end{align}
\end{subequations}
where
\begin{equation*}
J_\mathrm{ER}=k_f\mu([\ip])h\dfrac{bk_1+c}{k_1+c}\quad \mathrm{and} \quad J_\mathrm{pump}=\dfrac{\gamma c}{k_\gamma+c}\,.
\end{equation*}

Equations \eqref{nonlocal_model;a} and \eqref{nonlocal_model;b} are the spatially extended Atri model for \ca signalling. In equation \eqref{nonlocal_model;a} the term $J_\mathrm{ER}$ is the flux of \ca from the ER into the cytosol through \iprs, where the constant $k_f$ is the calcium flux when all \iprs are open and activated, $b$ is a basal current through the \iprs, and $\mu([\ip])=[\ip]/\left(k_\mu+[\ip]\right)$ is the fraction of the \iprs that have \ip bound and is an increasing function of $[\ip]$. In the spatially clamped Atri model relaxation oscillations can be sustained at constant $[\ip]$, and $\mu$ is a bifurcation parameter (see \cite{atri1993single, kaouri2019simple} for representative bifurcation diagrams). $J_\mathrm{pump}$ is the \ca flux through the SERCA pumps where $\gamma$ is the maximal pump rate and $k_\gamma$ is the \ca concentration at which the pump rate is at half-maximum. In equation (\ref{nonlocal_model;b}) the time constant $\tau_h>1 \s$ represents the slower time-scale of the inactivation of the \ipr by \ca compared to its activation \citep{atri1993single,dupont2016models}. Equations \eqref{nonlocal_model;c} is a non-local, non-linear PDE for the cell density that combines diffusion, cell-cell adhesion (advection) and proliferation (see \cite{domschke2014mathematical} and references therein). All parameter values can be found in Tables \ref{tab:1} and \ref{tab:2}.

\subsection{Effect of \ca on cell proliferation}
The role of \ca signals in the proliferation of cancer cells is cancer type specific due to differences in the behaviour of the \ca-conducting channels and pumps \citep{monteith2017calcium}. Here, we assume that \ca enhances the proliferation rate since it has been shown that \iprs are upregulated in cancer \citep{monteith2007calcium,monteith2017calcium}, leading to an enhanced proliferation and survival in all types of cancer \citep{cardenas2016selective,prevarskaya2018ion,rezuchova2019type,tsunoda2005inositol}. Moreover, assuming that cancer cells proliferate in a logistic manner (to describe the observed slow-down in tumour growth following the loss of nutrients \citep{laird1964dynamics}), we choose the growth function $f{\left(c,u\right)}$ as
\begin{equation}
f{\left(c,u\right)}=r_1{\biggl(1+g{\left(c\right)}\biggr)}\,u{\left(1-\dfrac{u}{k_u}\right)},
\label{prolif_term}
\end{equation}
where $r_1$ is the basal growth rate of $u$ and $k_u$ is the carrying capacity. The \ca-dependent function $g\left(c\right)$ describes the enhanced proliferation of cancer cells that is associated with a major re-arrangement of \ca pumps, \ce{Na^+}/\ca exchangers and \ca channels \citep{capiod2007calcium,simpson1986calcium,taylor1992inhibition}; we assume that it is given by
\begin{equation}
g\left(c\right)=\dfrac{r_2c^2}{r_3+c^2},
\label{hill_prolif}
\end{equation}
i.e. it saturates as $c$ increases and vanishes at $c=0$. We choose $r_1$, $r_2$ and $r_3$ based on experimental evidence, as follows. For $r_1$ it was shown that doubling times for cancer cells range from $1-10$ days \citep{cunningham2015vitro,morani2014pten}. In \cite{panetta2000modelling} the doubling time for breast and ovarian cancer ranges between $0.25-7$ days. Here we choose as doubling time $7$ days. $r_2$, the highest reaction rate that can be achieved at saturating \ca concentrations, and the half-maximal \ca concentration constant, $r_3$, are based on experimental evidence in \citep{simpson1986calcium,taylor1992inhibition}. All parameter values can be found in Tables \ref{tab:1} and \ref{tab:2}.

%where the effect of the increase of \todo{everywhere replace with }calcium concentration on breast cancer cell line HT-39 proliferation is measured; from these studies we see that a raising \ca from $1.3 \mu M$ to $2.6 \mu M$ stimulates cell proliferation of HT-39 cultures by $51.8\pm 2.3\%$. Thus, here we choose the values $r_2=1.6$ and $r_3=1.96\mu M^2$. \todo[inline]{VB: Should we mention here the value $r_3=1.96\mu M^2$ since we got the non-dim value of $r_3$ from this experiment?}

%This corresponds to dimensionless values of the growth rate in the range $\left(\ln\left(2\right)/10, \ln\left(2\right)/1\right)=\left(0.07, 0.7\right)$. In \cite{panetta2000modelling} the doubling time for breast and ovarian cancer ranges between $0.25-7$ days, which corresponds  to dimensionless values of growth rate between $0.1$ and $2.8$. Here we choose the dimensionless value $0.1$.

\subsection{Effect of \ca on cell-cell adhesion}
Cancer cells often show a decrease in cell-cell adhesion compared to healthy cells, which correlates with tumour invasion and metastasis \citep{cavallaro2001cell,makena2020subtype}. When adhesive bonds are formed and broken a cell-cell adhesion-mediated directed cancer cell migration occurs as a result of cellular  attraction and repulsion.  The cell-cell adhesion forces are created through the binding of adhesive molecules such as cadherins \citep{byers1995role,kim2011calcium, panorchan2006single}, see Section~\ref{intro}. Thus, we consider a calcium-dependent adhesion term in a bounded domain $\Omega=\left[0, R_s\right]$ in the cell density equation \eqref{nonlocal_model;c} where the non-local cell-cell interactions are described by a function that depends on cell density and \ca,
\begin{align*}
F\left[c,u\right]&=\dfrac{S\left(c\left(x,t\right)\right)}{R_s}\int_{0}^{R_s} K_\mathrm{int}\left(r\right)\biggl(u\left(x+r, t\right)-u\left(x-r, t\right)\biggr)\mathrm{d}r,
\end{align*}
where $K_\mathrm{int}\in L^\infty\left(\Omega\right)$ is the interaction kernel between cancer cells, with $\partial_x K_\mathrm{int}\in L^\infty\left(\Omega\right)$, and $S\left(c\right)$ the adhesion strength function, which depends on \ca. $R_s>0$ is the cell sensing `radius', i.e. the maximum range over which a cell can detect surrounding cells \citep{armstrong2006continuum}. Here, we assume that $R_s$ equals five times the length of an average cell \citep{armstrong2006continuum,gerisch2008mathematical}. Biologically this represents the extent of the cell's protrusions, e.g. filopodia. We define an attraction-repulsion kernel (see \citep{eftimie2007modeling, eftimie2017pattern}) as
\begin{equation*}
K_\mathrm{int}\left(x\right)=q_aK_{a}\left(x\right)-q_rK_r\left(x\right),
\end{equation*}
with $q_{a}$ and $q_{r}$ describing the magnitude of attractive and repulsive interactions, respectively, and $K_{a}(x)$ and $K_{r}(x)$ denoting the spatial range over which these interactions take place. We will take the kernel to be attractive at medium/long ranges (i.e.  at the edges of the cell) ensuring cell cohesion, and repulsive at very short ranges (i.e.  over the cell surface) to represent cell volume-exclusion effects and thus prevent unrealistically high cell densities \citep{palachanis2015particle}. Throughout the rest of this study, we consider Gaussian attraction and repulsion kernelw \citep{eftimie2007modeling} so that
\begin{equation}
K_\mathrm{int}\left(x\right)=\dfrac{q_a}{\sqrt{2\pi m^2_{a}}}e^{-\frac{\left(x-s_{a}\right)^2}{2m^2_{a}}}-\dfrac{q_r}{\sqrt{2\pi m^2_{r}}}e^{-\frac{\left(x-s_{r}\right)^2}{2m^2_{r}}},\label{KernelTranslateGauss}
\end{equation}
where $s_a$ and $s_r$ represent the location of maximal attraction and repulsion, respectively, with $s_r<s_a<R_s$. The constants $m_{j}=s_{j}/8, j=a, r$, represent the widths of the interaction kernels, respectively. They are chosen such that the support of more than $98\%$ of the mass of the kernels is inside the interval $\left[0, \infty\right)$ %\todo{AB: I'm very much confused by this sentence. Why do we care about the interval $[0, \infty)$ when we integrate over a compact set? -- VB: This is the definition of the model. Then we look at a specific region.} \citep{eftimie2007modeling,eftimie2017pattern}.
%\todo[inline]{AB: Generalise our approach by referring to Leah Keshet's kernel types you have mentioned.}

As discussed in Section \ref{intro}, expression of \ca-dependent cell-cell adhesion molecules is reduced in several human cancer types when \ca levels are increased \citep{byers1995role,cavallaro2004cell}, which leads to a decreased adhesive force between the cells. A biologically realistic choice for the adhesion strength function is thus
\begin{align}
S\left(c\right)&=s^{\star}\biggl(1-\dfrac{a_1c}{a_2+c}\biggr)\,
\label{strength_functions}
\end{align}
an inverse Hill function for $c$ that tends to zero for large $c$ values.
 We estimated the parameters $a_1, a_2$ and $s^{\star}$ so that the adhesive force exhibits a biologically sensible response to \ca variations (for parameter values see Table \ref{tab:2}).

 %\ca is a small molecule involved in rapid intracellular oscillations but the much larger cells are not expected to respond very prominently to \ca changes. \todo{AB/RT: please comment/refine this}%based on the range of the adhesion strength parameters in \cite{armstrong2006continuum,gerisch2008mathematical} (see also Table~\ref{tab:2}).

\subsection{Non-dimensionalized model}
To non-dimensionalize the model (\ref{nonlocal_model}), we define the following quantities:
\begin{align*}
\tilde{t}=\dfrac{t}{\tau_h}, \;\;\; \tilde{x}=\dfrac{x}{L_0}, \;\;\;  \tilde{c}=\dfrac{c}{k_1}, \;\;\;  &\tilde{u}=\dfrac{u}{k_u}, \;\;\; \tilde{R}_s=\dfrac{R_s}{L_0},\;\;\; \tilde{q}_{a}=k_uq_{a}, \;\;\;\tilde{q}_{r}=k_uq_{r},\\ &\tilde{S}(\tilde{c})=\dfrac{\tau_h}{L_0^2}S(k_1\tilde{c}).
\end{align*}
The length scale, $L_0$, is defined as the typical cell size/diameter of an average cancer cell.
Cancer cells can be smaller or bigger than healthy cells depending on several factor including the cancer type. HeLa cells, for example, are around $40\mm$ in diameter, while they measure $20\mm$ in their naturally compressed state \citep{boulter2006regulation,puck1956clonal}. Generally, the average cancer cell diameter is between $20-30\mm$ \citep{ha2011essentials}. Here, we choose $L_0=20\mm$, while we set the time scale as $\tau_h=2\s$ \citep{kaouri2019simple}. In addition, we rescale the  cell density with the cell carrying capacity, $k_u$, taken to be $\sim 6.7\cdot 10^7$cell/volume \citep{gerisch2008mathematical}. We obtain the dimensionless parameters:
\begin{align*}
\widetilde{D}_c=\dfrac{D_c\tau_h}{L_0^2}, \;\;\; K_1&=\dfrac{k_f\tau_h}{k_1}, \;\;\; \Gamma=\dfrac{\gamma\tau_h}{k_1}, \;\;\; K=\dfrac{k_\gamma}{k_1}, \;\;\; K_2=\dfrac{k_2}{k_1},\\& \widetilde{D}_u=\dfrac{D_u\tau_h}{L_0^2},  \;\;\; \tilde{r}_1=r_1\tau_h, \;\;\; \tilde{r}_3=\dfrac{r_3}{k_1^2}.
\end{align*}
%As in \citep{atri1993single} we set $K_2=1$ from now on.
We also briefly discuss the choice of the diffusion coefficients. It has been shown in \citep{allbritton1992range} that the diffusion coefficient of free cytosolic \ca is $2.23\cdot 10^{-6}\cms$. The action of omnipresent \ca buffers can be subsumed into an effective \ca diffusion coefficient, which we here set to $D_c=0.2\cdot 10^{-6}\cms$. Assuming that the delay of \ca propagation through gap junctions joining cells is negligible, we arrive at $\widetilde{D}_c=0.1$. The diffusion coefficient of cancer cells is in the range of $10^{-11}-10^{-9}\cms$ \citep{bray1992cell,chaplain2006mathematical, franssen2019mathematical}. This corresponds to dimensionless values of $\widetilde{D}_u$ between  $5\cdot 10^{-6}-5\cdot 10^{-3}$. We choose $\widetilde{D}_u=0.0025$. All other model parameters are given in  Tables~\ref{tab:1} and ~\ref{tab:2}.
%\begin{equation*}
%\tilde{D}_u=\dfrac{D_u\tau_h}{L_0^2}=\dfrac{0.5 \mu m^2 s^{-1} 2s}{400 \mu m^2}=\dfrac{1}{400}=0.0025.
%\end{equation*}

As in \citep{domschke2014mathematical}, we introduce the dimensionless functions  $\widetilde{K}_{a,r}\left(\widetilde{r}\right)=L_{0}K_{a,r}\left(L_0\tilde{r}\right)=L_{0}K_{a,r}\left(r\right)$ so that
\begin{equation*}
\widetilde{K}_\mathrm{int}\left(\tilde{r}\right)=L_{0}k_u\left(q_aK_{a}\left(r\right)-q_rK_{r}\left(r\right)\right)\,.
\end{equation*}
Therefore, we have for the non-local term
\begin{equation*}
\begin{split}
F&\left[c,u\right]\left(x,t\right)=\\
&=\dfrac{L_0}{\tau_h \tilde{R}_s}\widetilde{S}\left(\tilde{c}\right)\tilde{q}_a\int_0^{\tilde{R}_s} \biggl(\widetilde{K}_a\left(\tilde{r}\right)-\dfrac{\widetilde{q}_r}{\widetilde{q}_a}\widetilde{K}_r\left(\tilde{r}\right)\biggr)\left(\tilde{u}\left(\tilde{x}+\tilde{r},\tilde{t}\right)-\tilde{u}\left(\tilde{x}-\tilde{r},\tilde{t}\right)\right)\mathrm{d}r\\
&= \dfrac{L_0}{\tau_h\tilde{R}_s}\tilde{s}^{\star}\widetilde{q}_a \tilde{F}\left[\tilde{c}, \tilde{u}\right]\left(\tilde{x},\tilde{t}\right)=F_0\tilde{F}\left[\tilde{c}, \tilde{u}\right],
\end{split}
\end{equation*}
where $F_0=L_0 \tilde{s}^{\star}\tilde{q}_a/(\tau_h\tilde{R}_s)$ is the typical cancer cell speed.
%\begin{equation*}
%\tilde{F}\left[\tilde{c}, \tilde{u}\right]:=\dfrac{1}{F_0} F\left[c, u\right].
%\end{equation*}

\cite{clark2015modes} showed that cancer cell speeds cannot exceed $10\mm/\mathrm{min}$. We Consider the typical cancer cell speed, $F_0$, to vary between $1\mm/\mathrm{min}$ and $10\mm/\mathrm{min}$ to account for various cancer types which are characterised by slower or faster cells (e.g. for A375M2 human melanoma the speed ranges between $0.5 -- 10 \mm/\mathrm{min}$, and for MDA-MB-231 breast cancer it ranges between $0.4 -- 4.2 \mm/\mathrm{min}$). We find that the ratio $\tau_h F_0/L_0$ is in the range
\begin{equation*}
0.0017\leq \dfrac{\tau_h}{L_0}F_0 \leq0.017,
\end{equation*}
leading to
\begin{equation}
0.008\leq \tilde{s}^{\star}\tilde{q}_a\leq 0.08.
\label{admiss_values_qas}
\end{equation}
This provides bounds for the value of $\tilde{s}^{\star}\tilde{q}_a$ we are going to choose in Sections \ref{analytical_results} and \ref{numerics}.
%\todo[inline]{Andreas: Can you add a sentence here regarding what you told us re friction coeffcients and their wide variation and hence this bound being ok?}
%\textbf{KK: perhaps write something along these lines? }The magnitudes of attraction and repulsion, $q_a$ and $q_r$, respectively, were estimated so that the speed of the movement of the cancer cells agrees most of the times with experimental evidence (see e.g. \cite{clark2015modes}). These values were also chosen in \cite{bitsouni2017mathematical,bitsouni2018aggregation}.

After dropping the tildes for notational convenience, we obtain the following non-dimensional system:
\begin{subequations}
\label{Nondim_AtriCellsDiff_nonlocal}
\begin{align}
\dfrac{\partial c}{\partial t}&=D_c\dfrac{\partial^2 c}{\partial x^2}+\mu K_1 h\dfrac{b+c}{1+c}-\dfrac{\Gamma c}{K+c},\label{Nondim_AtriCellsDiff_nonlocal;a}\\
\dfrac{\partial h}{\partial t}&=\dfrac{1}{1+c^2}-h,\label{Nondim_AtriCellsDiff_nonlocal;b}\\
\dfrac{\partial u}{\partial t}&=D_u\dfrac{\partial^2 u}{\partial x^2}-\dfrac{\tau_h}{L_0}F_0\dfrac{\partial}{\partial x}\left(u\left(F[c,u]\right)\right)+r_1\left(1+\dfrac{r_2c^2}{r_3+c^2}\right)u\left(1-u\right).\label{Nondim_AtriCellsDiff_nonlocal;c}
\end{align}
\end{subequations}
Although $D_u=0.0025$, which corresponds to a large diffusion value, the behaviour of cancer cells is still advection-dominated. This directed, advective movement of cancer cells results from the aforementioned cell-cell adhesion forces and from an elevated
macrophage density near highly mutated cancer cells \citep{lin2006macrophages}, which decreases the random movement of the cancer cells \citep{goswami2005macrophages, hagemann2005macrophages}.

%%%%%%%%%%%%%%%%%%%-------- Table 1 ---------%%%%%%%%%%%%%%%%%%%%%%
\begin{longtable}{|p{.9cm}|p{5.5cm}|p{2.5cm}|p{1.5cm}|p{3cm}|}
\caption{Model parameters, dimensional values, non-dimensional values and relevant references.}\\
\hline
Param. & Description  & Dim. value & Non-dim. value  & Reference \\
\hline
\endfirsthead
\multicolumn{4}{c}
{\tablename\ \thetable\ -- \textit{Continued from previous page}} \\
\hline
Param.& Description  & Dim. value & Non-dim. value  & Reference \\
\hline
\endhead
\hline \multicolumn{4}{r}{\textit{Continued on next page}} \\
\endfoot
\hline
\endlastfoot
$D_c$ & Diffusion coefficient of \ca & $20\mms$ & $0.1$ & \cite{atri1993single,hofer2001intercellular,wilkins1998intercellular}  \\\hline
$b$ & Fraction of activated \iprs receptors when [\ca]=0  & - & 0.111 & \cite{atri1993single}\\\hline
$k_1$ &  $K_m$ (Michaelis constant) for activation of \iprs receptors by \ca & $0.7 \mM$  & 1 & \cite{atri1993single, kaouri2019simple}\\\hline
$k_f$ & \ca flux when all \iprs receptors are open and activated & $16.2\mM\is$ &  $K_1=324/7$ & \cite{atri1993single, kaouri2019simple}\\\hline
$k_\mu$ & $K_m$ (Michaelis constant)   for binding of \ip to its receptor & $0.7\mM$ &  1 &  \cite{atri1993single, kaouri2019simple}\\\hline
$\gamma$ & Maximum rate of pumping of ER \ca & $2 \mM\is$ & $\Gamma=40/7$ &  \cite{atri1993single}\\\hline
$k_\gamma$ & $[\ca]_c$ at which the rate of \ca pumping from the cytosol is at half-maximum & $0.1\mM$ &  $K=1/7$ &  \cite{atri1993single, kaouri2019simple}\\\hline
$k_2$ & $K_m$ (Michaelis constant) for inactivation of \ip receptors by \ca  & $0.7\mM$  & $K_2=1$ & \cite{atri1993single, kaouri2019simple}\\\thickhline
$D_u$ & Diffusion coefficient of cancer cells & $0.5\mms$ & $0.0025$ & \cite{bray1992cell,chaplain2006mathematical,enderling2006mathematical,franssen2019mathematical}\\\hline
$R_s$ & Sensing radius & $100\mm$ & $5$ &   \cite{armstrong2006continuum,gerisch2008mathematical}\\\hline
$k_u$ & Carrying capacity of the cancer cell population & $ 6.7\cdot 10^7\mathrm{cells}/\mathrm{cm}^3$ & 1 & \cite{gerisch2008mathematical}\\\hline
$r_1$ & Growth rate of the cancer cell population & $7$ days (doubling time) & $0.1$ &  \cite{cunningham2015vitro,morani2014pten,panetta2000modelling}
\label{tab:1}
\end{longtable}
%%%%%%%%%%%%%%%%%%%%%%%%%%%%%%%%%%%%%%%%%%%%%%%%%%%%%%%%%%%%%%%%%%%%%%%%%%%%%%%%%%%%%%%%%%%%

%%%%%%%%%%%%%%%%%%%-------- Table 2 ---------%%%%%%%%%%%%%%%%%%%%%%
\begin{longtable}{|p{.9cm}|p{8cm}|p{1.5cm}|p{3.5cm}|}
\caption{Estimated model parameters, non-dimensional values and relevant references.}\\
\hline
Param. & Description  & Non-dim. value  & Reference \\
\hline
\endfirsthead
\multicolumn{4}{c}
{\tablename\ \thetable\ -- \textit{Continued from previous page}} \\
\hline
Param.& Description  & Non-dim. value  & Reference \\
\hline
\endhead
\hline \multicolumn{4}{r}{\textit{Continued on next page}} \\
\endfoot
\hline
\endlastfoot
$q_a$ & Magnitude of attraction &  $0-0.44$  & Guided by linear stability analysis (Section \ref{lin_stability}) and the range \eqref{admiss_values_qas},  based on  \cite{clark2015modes}\\\hline
$q_r$ & Magnitude of repulsion &  $0-0.44$  & Guided by linear stability analysis (Section \ref{lin_stability})\\\hline
$s_a$ & Attraction range  & $1$ & \cite{bitsouni2018non} \\\hline
$s_r$ & Repulsion range & $0.25$ & \cite{bitsouni2017mathematical,bitsouni2018aggregation,bitsouni2018non} \\\hline
$m_a$ & Width of attraction kernel  & $1/8$ & \cite{bitsouni2018non}\\\hline
$m_r$ & Width of repulsion kernel  & $1/32$ & \cite{bitsouni2017mathematical,bitsouni2018aggregation,bitsouni2018non} \\\hline
$s^{\star}$ & Magnitude of cell-cell adhesion forces of the cancer cell population  & $1$ & \cite{armstrong2006continuum,bitsouni2017mathematical,bitsouni2018aggregation,gerisch2008mathematical}\\\hline
$a_1$ & Lowest value of cell-cell strength due to increase in [\ca]   & $0.5$ & Estimated\\ \hline
$a_2$ & Half-minimum ($K_m$) [\ca] & $0.5$ & Estimated\\ \hline
$r_2$ & Largest reaction value at saturating [\ca] & $1.6$ & \cite{simpson1986calcium,taylor1992inhibition}\\\hline
$r_3$ & Half-maximal [\ca] & $4$ & \cite{simpson1986calcium,taylor1992inhibition}
\label{tab:2}
\end{longtable}
%%%%%%%%%%%%%%%%%%%%%%%%%%%%%%%%%%%%%%%%%%%%%%%%%%%%%%%%%%%%%%%%%%%%%%%%%%%%%%%%%%%%%%%%%%%%
%%%%%%%%%%%%%%%%%%%%%%%%%%%%%%%%%%%%%%%%%%%%%%%%%%%%%%%%%%%%%%%%%%%%%%%%%%%%%%%%%%%%%%%%%%%%

\section{Analytical results}
\label{analytical_results}
\subsection{Existence of solution}
\label{Existence}
The existence of a unique global-in-time classical solution of the model (\ref{nonlocal_model}) can be proven using the theory of semigroups \citep{henry1981geometric}, within the framework of ODEs. The proof of the theorem follows the same steps as \citep{bitsouni2017mathematical,chaplain2011mathematical}.

%%%%%%%%%%%%%%%%%%%%%%%%%%%%%%%%%%%%%%%%%%%%%%%%%%%%%%%%%%%%%%%%%%%%%%%

%%%%%%%%%%%%%%%%%%%%%%%%%%%%%%%%%%%%%%%%%%%%%%%%%%%%%%%%%%%%%%%%%%%%%%%

\subsection{Linear stability analysis}
\label{lin_stability}
In our model an instability of a spatially homogeneous state can arise when advection effects increase; we will call this an advection-driven instability (ADI). The loss of stability leads to spatial patterns, which biologically correspond to cell aggregations \citep{keller1970initiation}.

%The collective behaviour of cancer cells that appear in aggregations simultaneously integrates cell movement and remodelling of tissue structures while patterns become established and remain intact within the group  \citep{friedl2004collective}. Thus, this tendency of cancer cells for aggregation increases the risk of invasion and can lead to cancer spreading (metastasis) by group migration \citep{knutsdottir2014mathematical}. Hence, aggregations are one of the key factors in cancer invasion and metastasis \citep{glinsky2003intravascular}.

In this section, we linearise the model~\eqref{Nondim_AtriCellsDiff_nonlocal} and investigate the conditions for ADIs. The spatially homogeneous steady states $\left(c^{\star},h^{\star}, u^{\star}\right)$ of the system \eqref{Nondim_AtriCellsDiff_nonlocal} are given by
\begin{equation}
\left(c^{\star}, \dfrac{1}{1+{c^{\star}}^2}, 0\right) \;\; \textnormal{and}\;\;  \left(c^{\star}, \dfrac{1}{1+{c^{\star}}^2}, 1\right),
\label{steady_states}
\end{equation}
with $c^{\star}\geq 0$ determined by the solution of the quartic equation
\begin{equation*}
c^{{\star}^4}+c^{{\star}^3}+\biggl(1-\mu\dfrac{K_1}{\Gamma}\biggr)c^{{\star}^2}+\biggl(1-\mu\dfrac{K_1}{\Gamma}\biggl(K+b\biggr)\biggr)c^{\star}-\mu\dfrac{K_1}{\Gamma}K b=0.
\label{relation_c_mu}
\end{equation*}
We seek conditions for a steady state $(c^{\star},h^{\star},u^{\star})$ to become unstable due to ADI. We thus consider small perturbations to the steady state, $\left(\bar{c}, \bar{h}, \bar{u}\right)$, such that $c\left(x,t\right)=c^{\star}+\bar{c}\left(x,t\right), h\left(x,t\right)=h^{\star}+\bar{h}\left(x,t\right), u\left(x,t\right)=u^{\star}+\bar{u}\left(x,t\right)$. Substituting these into~\eqref{Nondim_AtriCellsDiff_nonlocal}, linearising around the spatially uniform steady state, and using the notation $\bar y=(\bar c, \bar h, \bar u)$, we obtain
\begin{equation*}
    \PD{\bar y}{t}=D \PD{^2 \bar y}{x^2}-\PD{J_a}{x} +J_r \bar y\,,
\end{equation*}
where $D$ is a diagonal matrix with entries $(D_c,0,D_u)$ and $J_a=(0,0,\alpha)$, with
\begin{equation*}
    \alpha=\dfrac{u^{\star}}{R_s}S(c^{\star})\int_0^{R_s} K_\mathrm{int}(r)\left(\bar{u}(x+r,t)-\bar{u}(x-r,t)\right)\mathrm{d}r\,,
\end{equation*}
and
\begin{equation*}
J_r=
    \begin{bmatrix}
J_2(c^{\star},h^{\star}) & \begin{matrix} 0 \\ 0 \end{matrix} \\
 \begin{matrix} r_1u^{\star}\left(1-u^{\star}\right)\dfrac{2r_2r_3c^{\star}}{r_3+c^{{\star}^2}}\;  & \;0 \; \end{matrix} & \; r_1\left(1-2u^{\star}\right)\left(1+\dfrac{r_2c^{{\star}^2)}}{r_3+c^{{\star}^2}}\right)
\end{bmatrix}\,,
\end{equation*}
where
\begin{equation*}
    J_2=
    \begin{bmatrix}
\mu K_1 h \dfrac{1-b}{\left(1+c^{\star}\right)^2}-\dfrac{\Gamma K}{\left(K+c^{\star}\right)^2} \; & \; \mu K_1 \dfrac{b+c^{\star}}{1+c^{\star}}\\
-\dfrac{2c^{\star}}{\left(1+c^{{\star}^2}\right)^2} \; & \; -1
\end{bmatrix},
\end{equation*}
is the Jacobian of the linearised Atri model. We seek solutions of the form $\bar y=w e^{i\xi x+\lambda t}$, where $w=(A_c,A_h,A_u)$ with $\lvert A_c\rvert,\lvert A_h\rvert, \lvert A_u\rvert \ll 1$. The wave number and frequency of the perturbations are denoted by $\xi$ and $\lambda$, respectively. We then find
\begin{equation*}
    \lambda w=\left(J_d + J_r \right) w\,,
\end{equation*}
with
\begin{equation*}
J_d=\begin{bmatrix}
-D_c\xi^2 & 0 & 0\\
0 & 0 & 0\\
0 & 0 & -D_u\xi^2+2\xi u^{\star}\widehat{K}^s_\mathrm{int}(\xi)S(c^{\star})/R_s\,,
\end{bmatrix}
\end{equation*}
where $\widehat{K}_\mathrm{int}^s(\xi)=\int_0^{R_s} K_\mathrm{int}(r)\sin(\xi r)\mathrm{d}r$ is the Fourier sine transform of  $K_\mathrm{int}(r)$.

\begin{figure*}[!htbp]
\subfloat[\label{subfig1a}]{%
       \includegraphics[width=.48\textwidth]{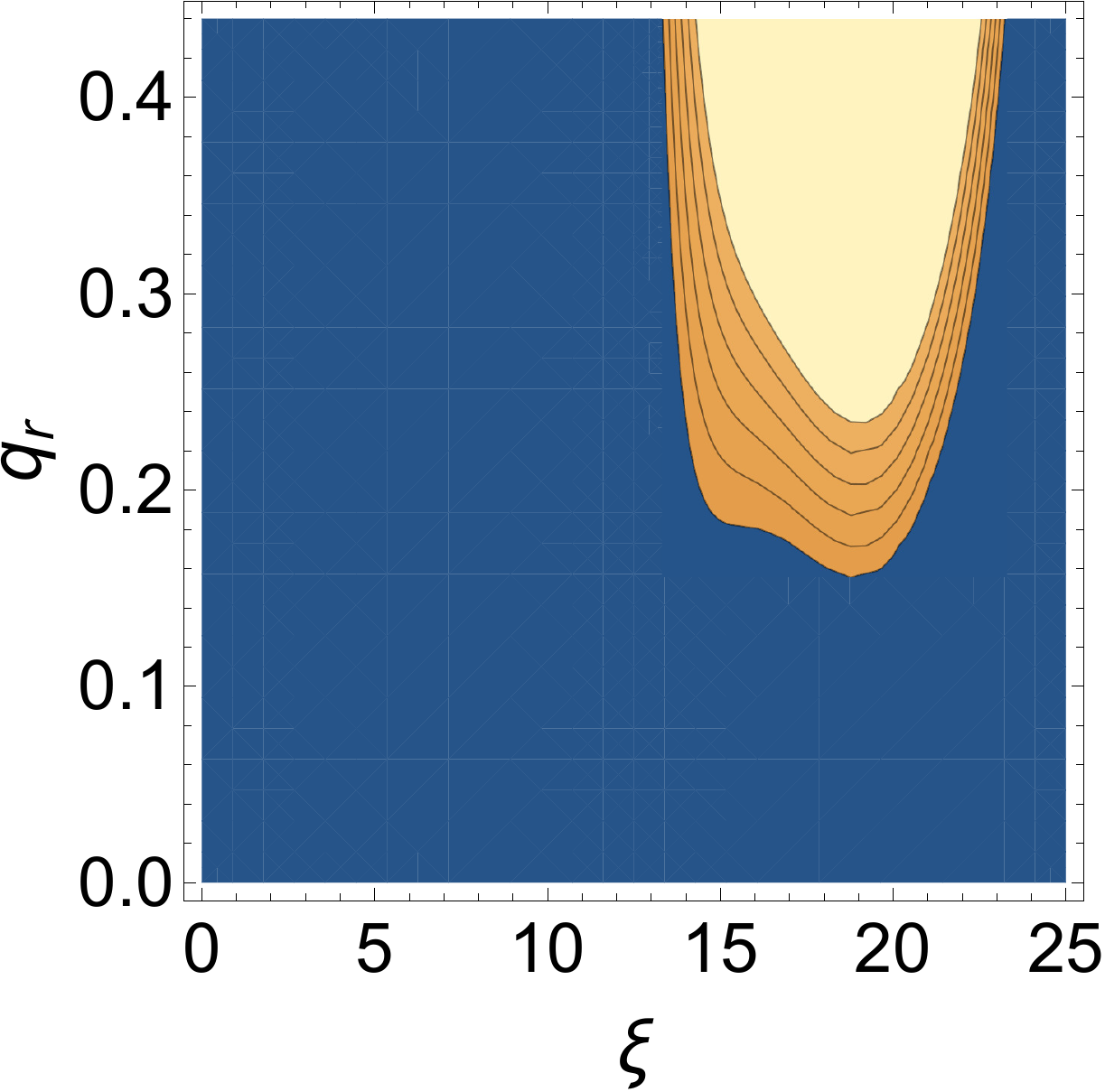}%{ContourPlotqa009qr}
     }
     \hfill
     \subfloat[\label{subfig1b}]{%
       \includegraphics[width=.48\textwidth]{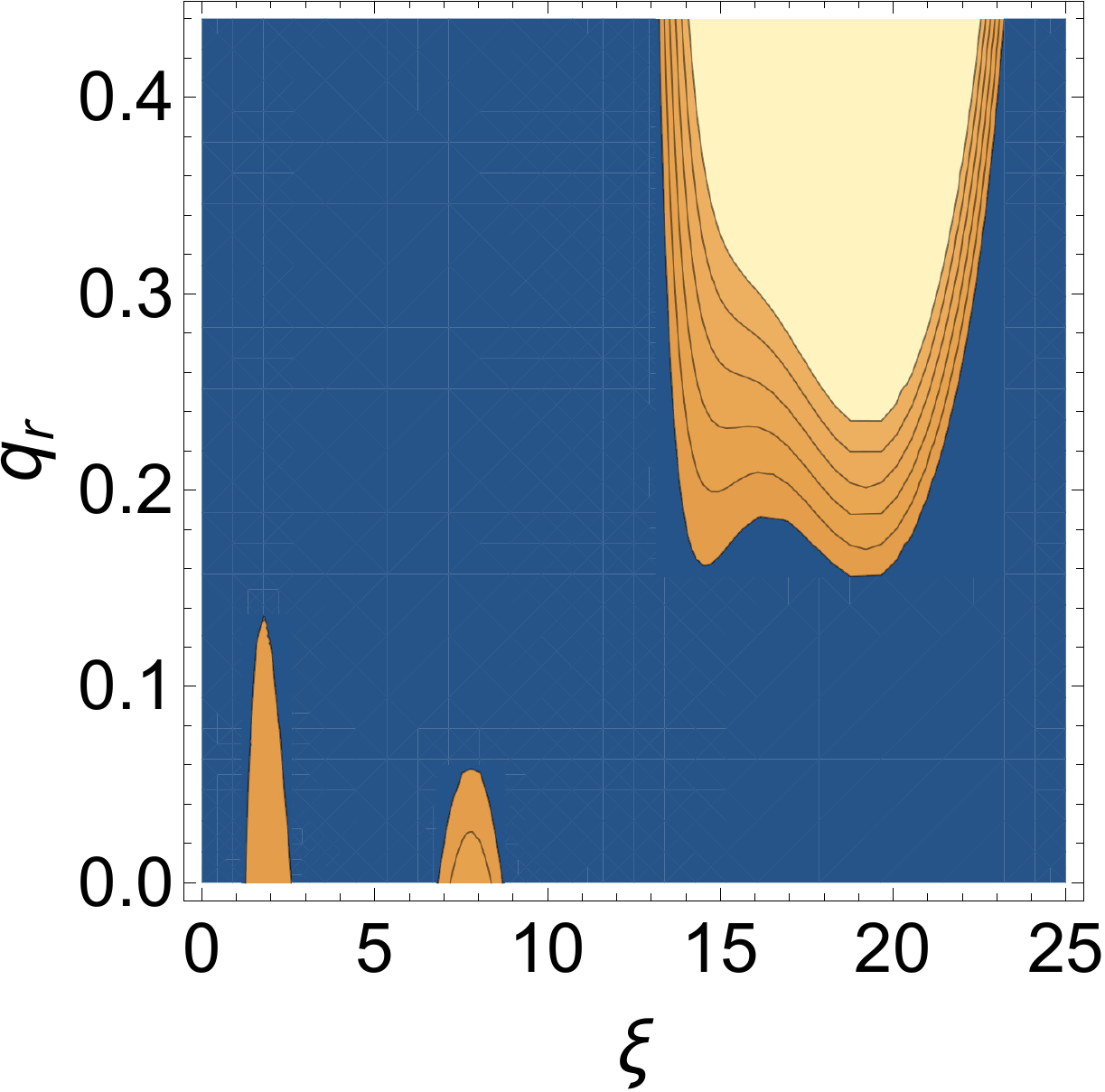}
     }\\
     \hfill
     \subfloat[\label{subfig1c}]{%
       \includegraphics[width=.48\textwidth]{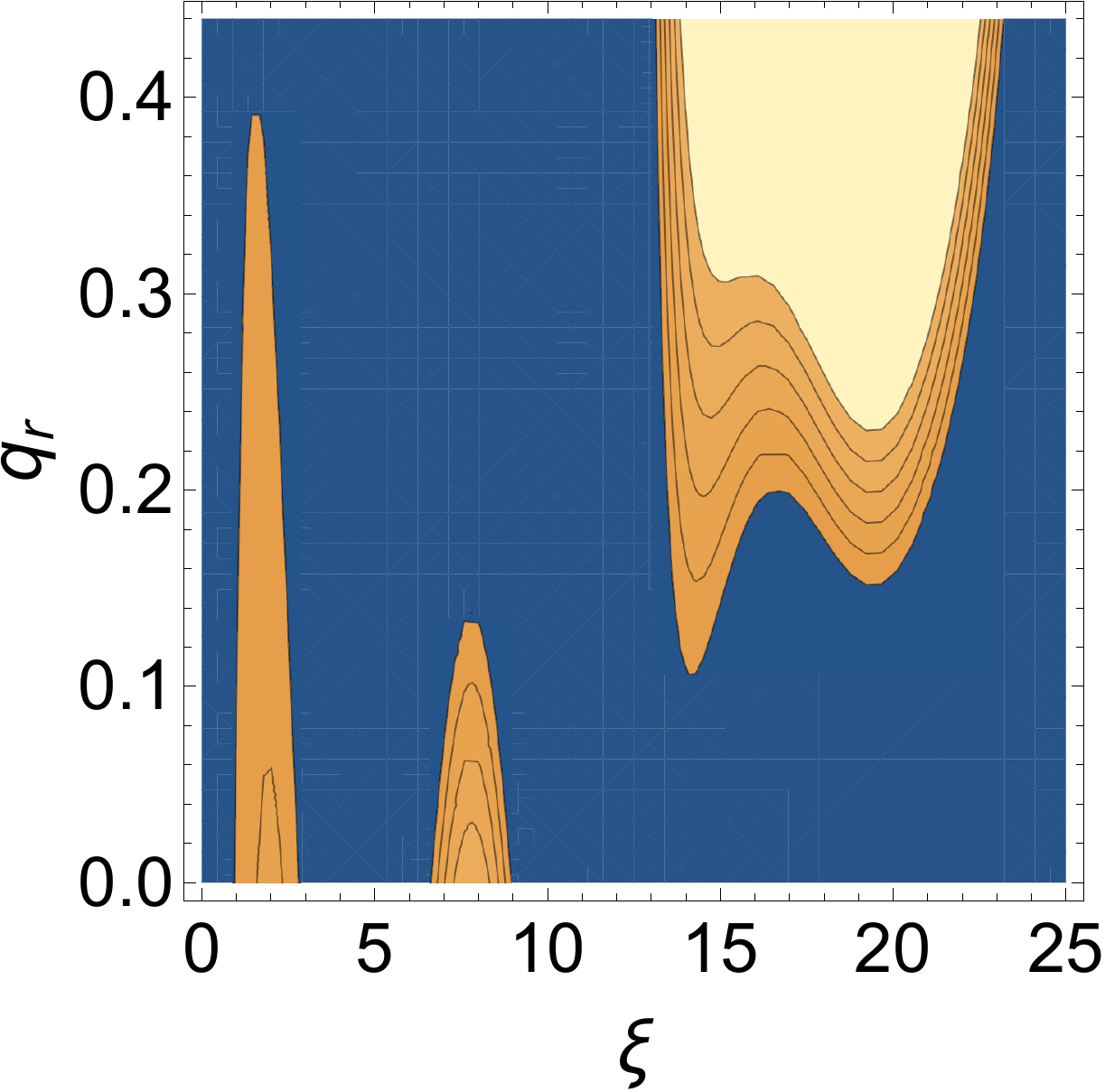}%{ContourPlotqr022qa}%
          }
            \hfill
     \subfloat[\label{subfig1d}]{%
       \includegraphics[width=.48\textwidth]{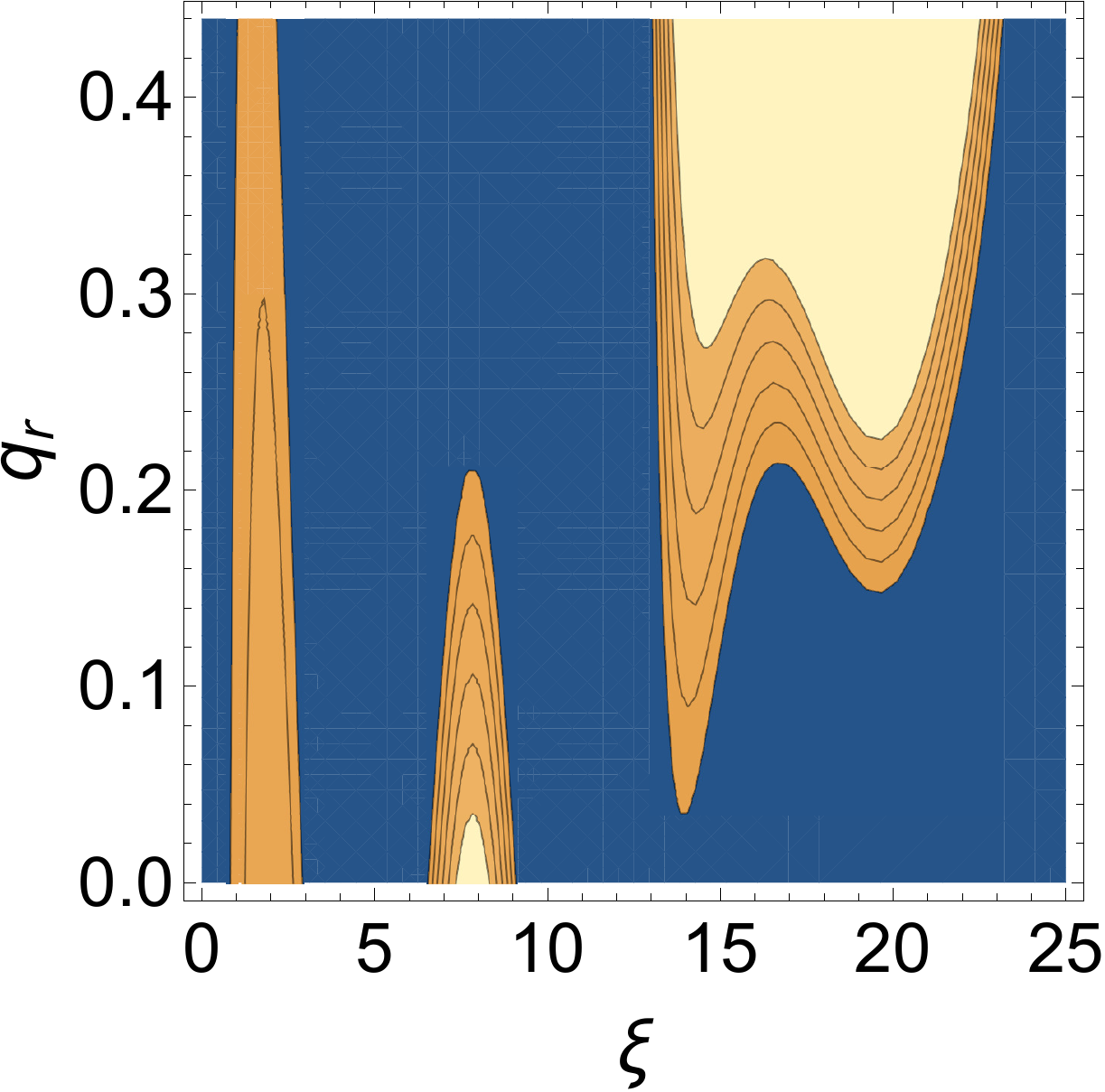}%{ContourPlotqr022qa}
          }
\caption{The contours of non-negative $\lambda_u^1(\xi,0)$, dispersion relation of the linearised cell density equation, for $c^{\star}=0$, $u^{\star}=1$, which enclose parameter regions corresponding to adhesion-driven instabilities, for: (a) $q_a=0.14$ (b) $q_a=0.22$ (c) $q_a=0.33$ (d) $q_a=0.44$. In (a)--(d) $q_r$ varies from $0$ to $0.44$, respectively. The remaining parameter values are given in  Tables~\ref{tab:1} and ~\ref{tab:2}. Negative values of $\lambda_u^1(\xi,0)$ have been set to zero for better visualisation.}
\label{Fig_1}.
\end{figure*}

Since the cell density equation \eqref{Nondim_AtriCellsDiff_nonlocal;c} is not coupled to the  Atri equations \eqref{Nondim_AtriCellsDiff_nonlocal;a} and \eqref{Nondim_AtriCellsDiff_nonlocal;b}, the matrix $M=J_d+J_r$ has a block structure and the eigenvalues of $M$ are split into those of the (linearised) Atri model and that of the (linearised) cancer cell density equation. Hence, to identify ADIs we only need to study the linear stability of the cell density equation, i.e. the eigenvalue (dispersion relation)
\begin{equation*}
\lambda_u\left(\xi,c^{\star}\right)=-D_u\xi^2+\dfrac{2\xi u^{\star}}{R}S(c^{\star})\hat{K}^s_\mathrm{int}(\xi)+r_1\left(1-2u^{\star}\right)\left(1+\dfrac{r_2{c^{\star}}^2}{r_3+{c^{\star}}^2}\right),
\end{equation*}
which for the Gaussian attraction and repulsion kernels given in (\ref{KernelTranslateGauss}) becomes
\begin{align}
\lambda_u\left(\xi,c^{\star}\right)=&-D_u\xi^2+\dfrac{2\xi u^{\star}}{R}S(c^{\star})\left(e^{-\frac{\left(\xi m_a\right)^2}{2}}\sin(\xi s_a)-\dfrac{q_r}{q_a} e^{-\frac{\left(\xi m_r\right)^2}{2}}\sin(\xi s_r)\right)\nonumber\\&\hspace{3.5cm}+r_1\left(1-2u^{\star}\right)\left(1+\dfrac{r_2{c^{\star}}^2}{r_3+{c^{\star}}^2}\right).
\label{lambda1_ss}
\end{align}
Solutions with $\lambda_u>0$ are unstable and grow exponentially in time, corresponding to pattern formation and cell aggregation in the non-linear system \citep{murray2003ii,painter2015nonlocal}.
For $u^{\star}=0$ and $\xi=0$, we obtain
\begin{equation*}
    \lambda_u^0(0,c^{\star})=r_1 \left(1+\dfrac{r_2{c^{\star}}^2}{r_3+{c^{\star}}^2}\right) > 0\,,
\end{equation*}
%so that $\lambda_u(\xi)$ remains positive for some $\xi>0$.
In contrast, for $u^{\star}=1$ and $\xi=0$, we find
\begin{equation*}
    \lambda_u^1(0,c^{\star})=-r_1 \left(1+\dfrac{r_2{c^{\star}}^2}{r_3+{c^{\star}}^2}\right) < 0.\\
\end{equation*}
Here, we use the superscript to indicate the value of $u^{\star}$. Note that $\lambda_u^0(\xi,0)$ and $\lambda_u^1(\xi,0)$ are the eigenvalues of the linearised Fisher's equation when $q_a=q_r=0$ . For positive $q_a$ and $q_r$ and no calcium,
$\lambda_u^1(\xi,0)$ becomes positive for some $\xi>0$ when the advection strength increases sufficiently. To identify the threshold values of $q_a$ and $q_r$, we present the non-negative contour plots of $\lambda_u^1(\xi,0)$ in Fig.~\ref{Fig_1}, where negative values are mapped to zero for better visualisation. In Figs.~\subref*{subfig1a}--\subref*{subfig1d} we set $q_a=0.14$, $q_a=0.22$, $q_a=0.33$ and $q_a=0.44$, respectively, while $q_r$ varies from $0$ to $0.44$. We observe extended regions with $\lambda^1_u>0$, which indicate pattern formation in the nonlinear system via ADI. In Figs.~\subref*{subfig1b}--\subref*{subfig1d} we observe disjoint parameter regions, which grow larger as $q_a$ increases.
%%%%%%%%%%%%%%%%%%%--------Fig 1 ---------%%%%%%%%%%%%%%%%%%%%%%
Note that we do not need to plot for larger values of $\xi$ since $\lambda^1_u(\xi,0)$ tends to $-\infty$ as $\xi$ tends to $\infty$.
%%%%%%%%%%%%%%%%%%%%%%%%%%%%%%%%%%%%%%%%%%%%%%%%%%%%%%%%%%%%%%%%%%%%%%%%%%%%%%%%%%%%%%%%%%%%

\begin{figure}[!htbp]
     \subfloat[\label{subfig2a}]{%
       \includegraphics[width=.3\textwidth]{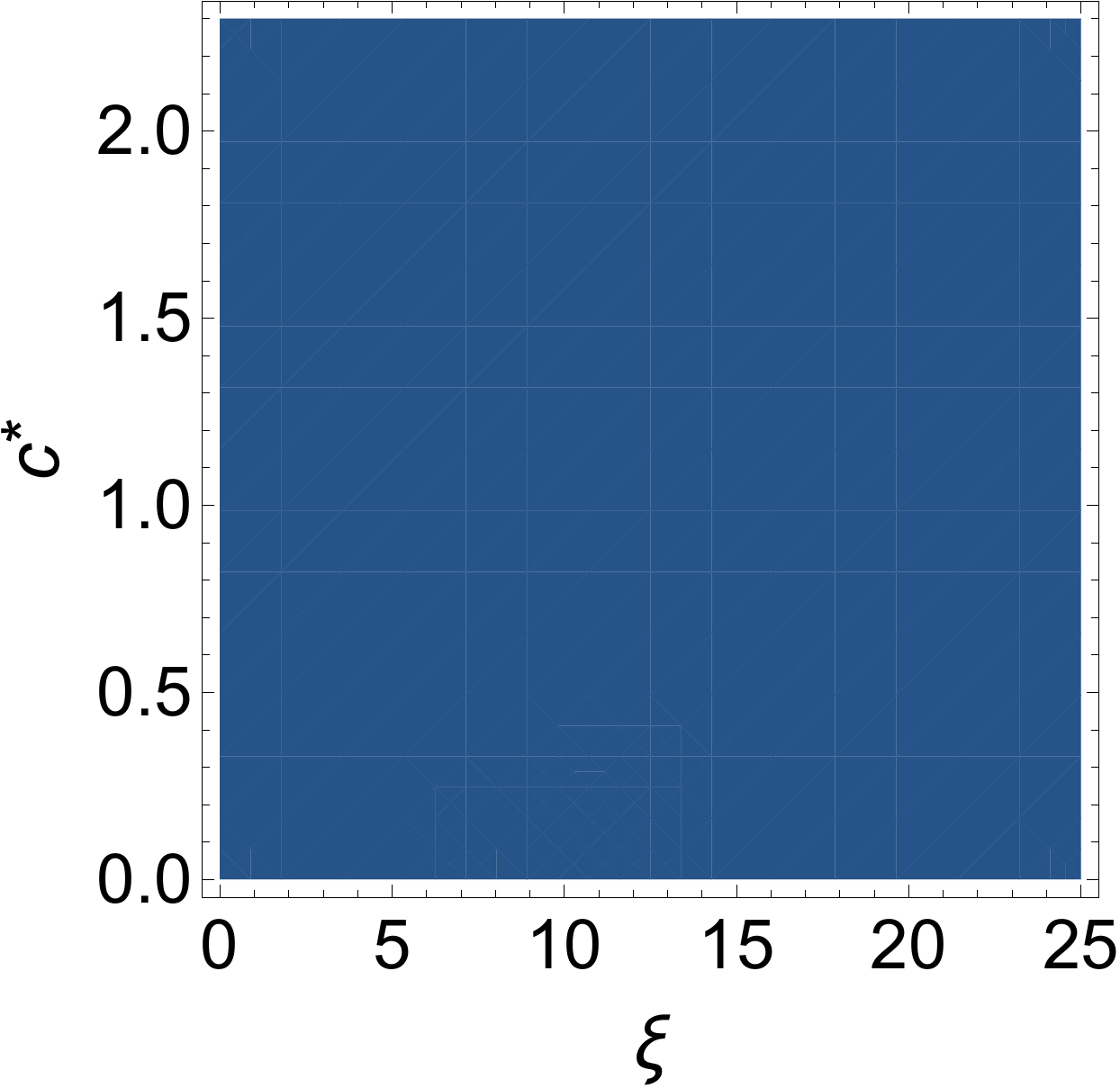}%
     }
     \hfill
   \subfloat[\label{subfig2b}]{%
       \includegraphics[width=.3\textwidth]{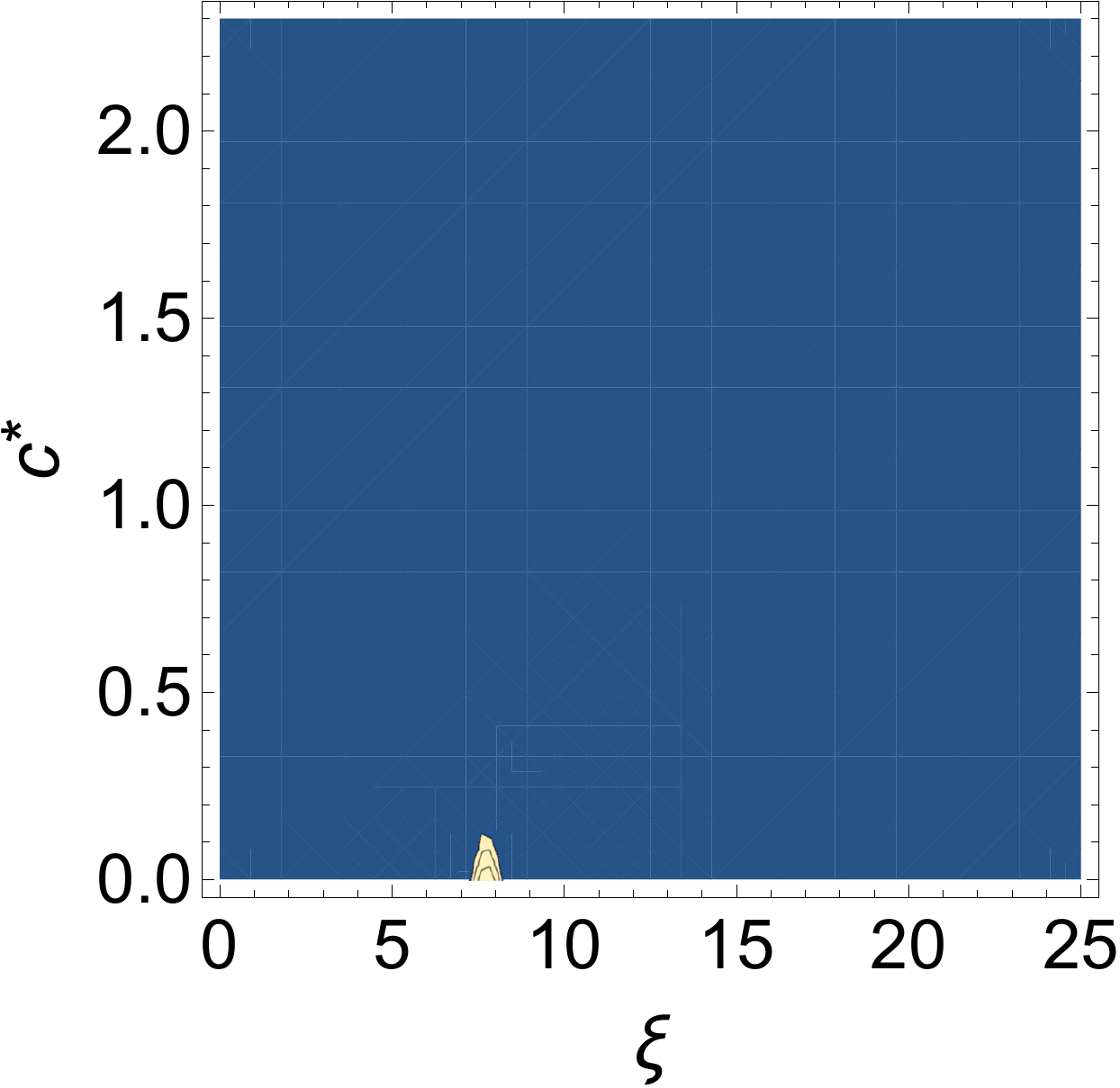}%
     }
     \hfill
     \subfloat[\label{subfig2c}]{%
       \includegraphics[width=.3\textwidth]{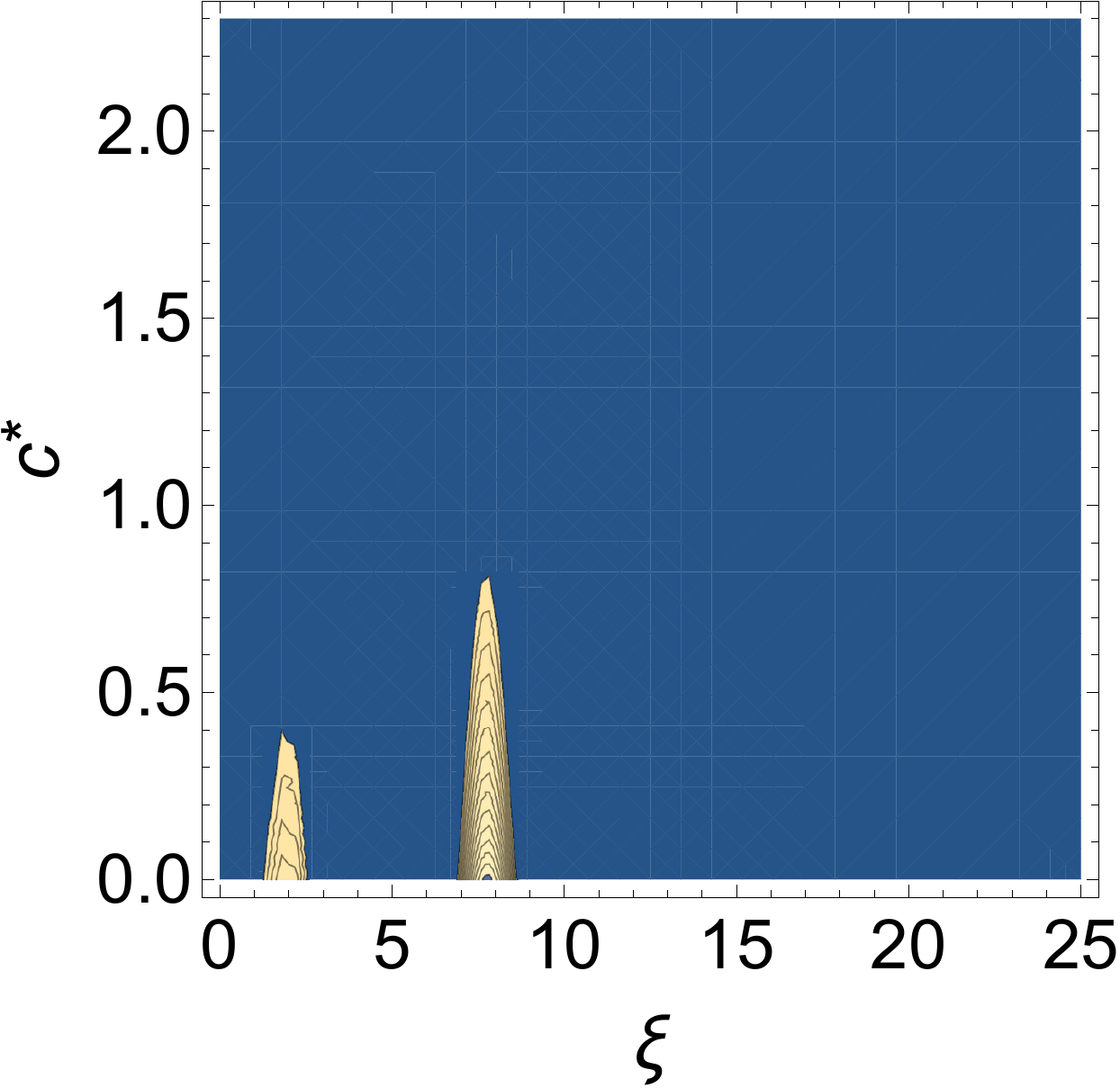}%
     }\\
      \subfloat[\label{subfig2d}]{%
       \includegraphics[width=.3\textwidth]{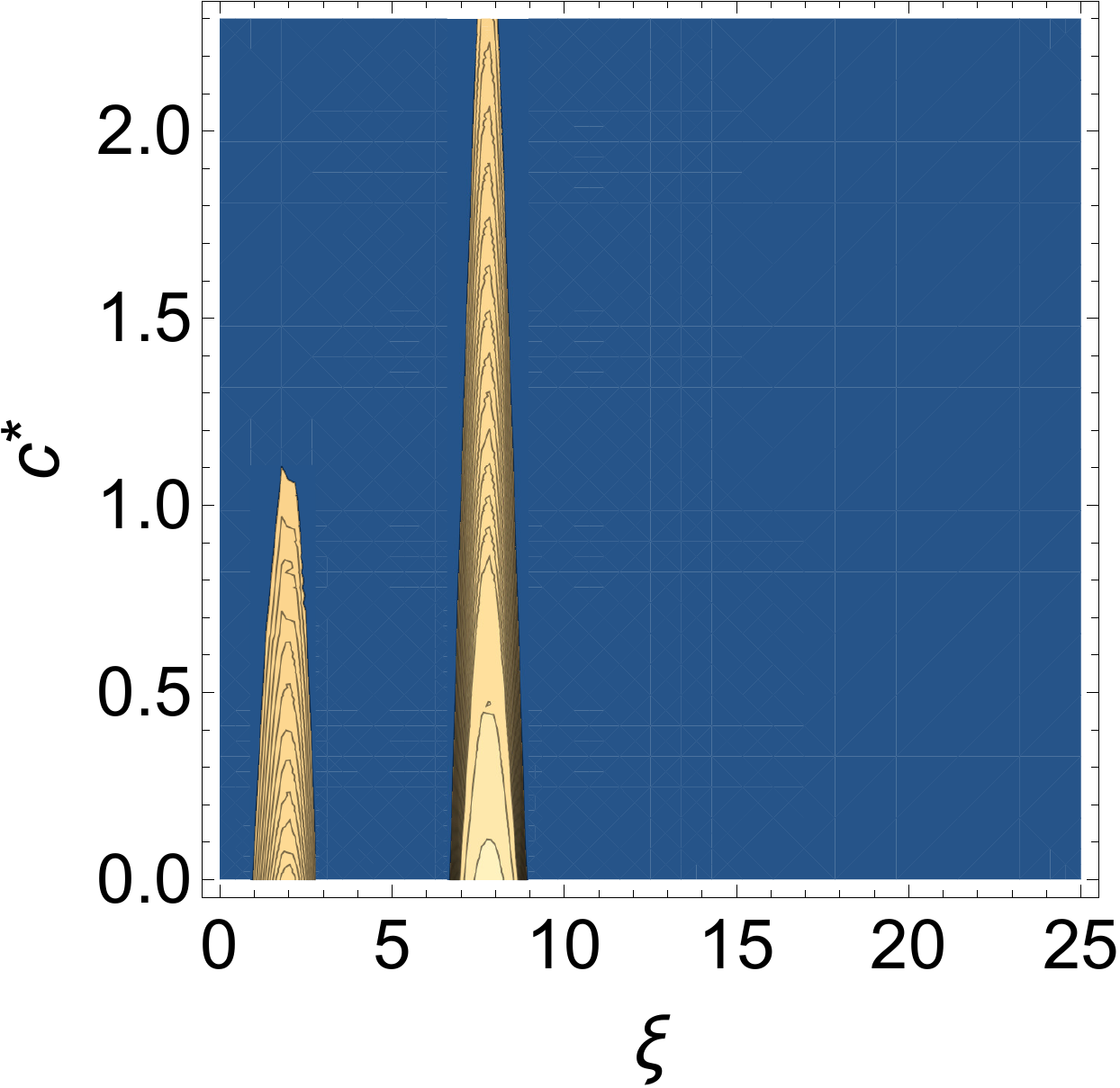}%
     }\hfill
     \subfloat[\label{subfig2e}]{%
       \includegraphics[width=.3\textwidth]{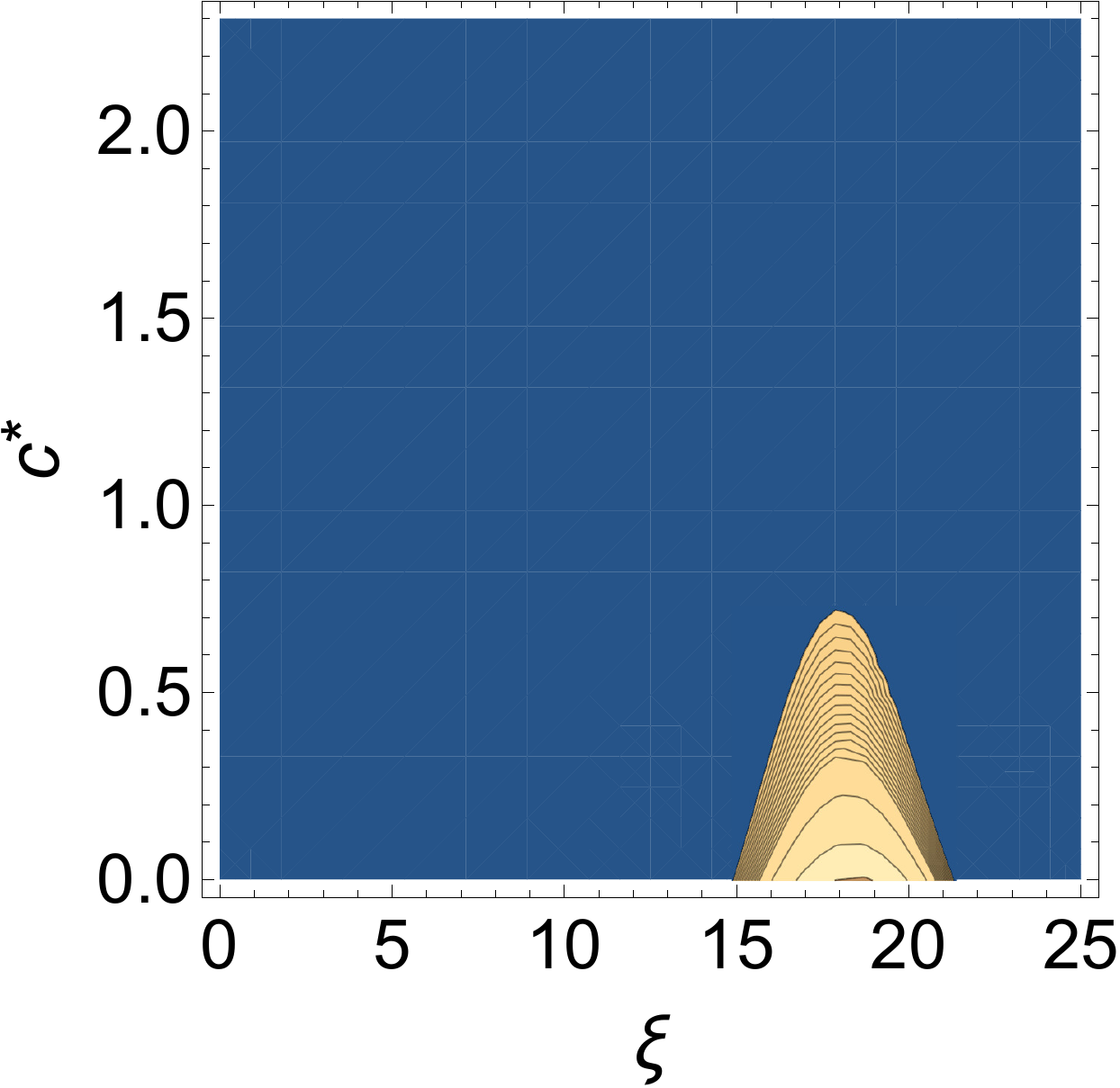}%
     }\hfill
          \subfloat[\label{subfig2f}]{%
       \includegraphics[width=.3\textwidth]{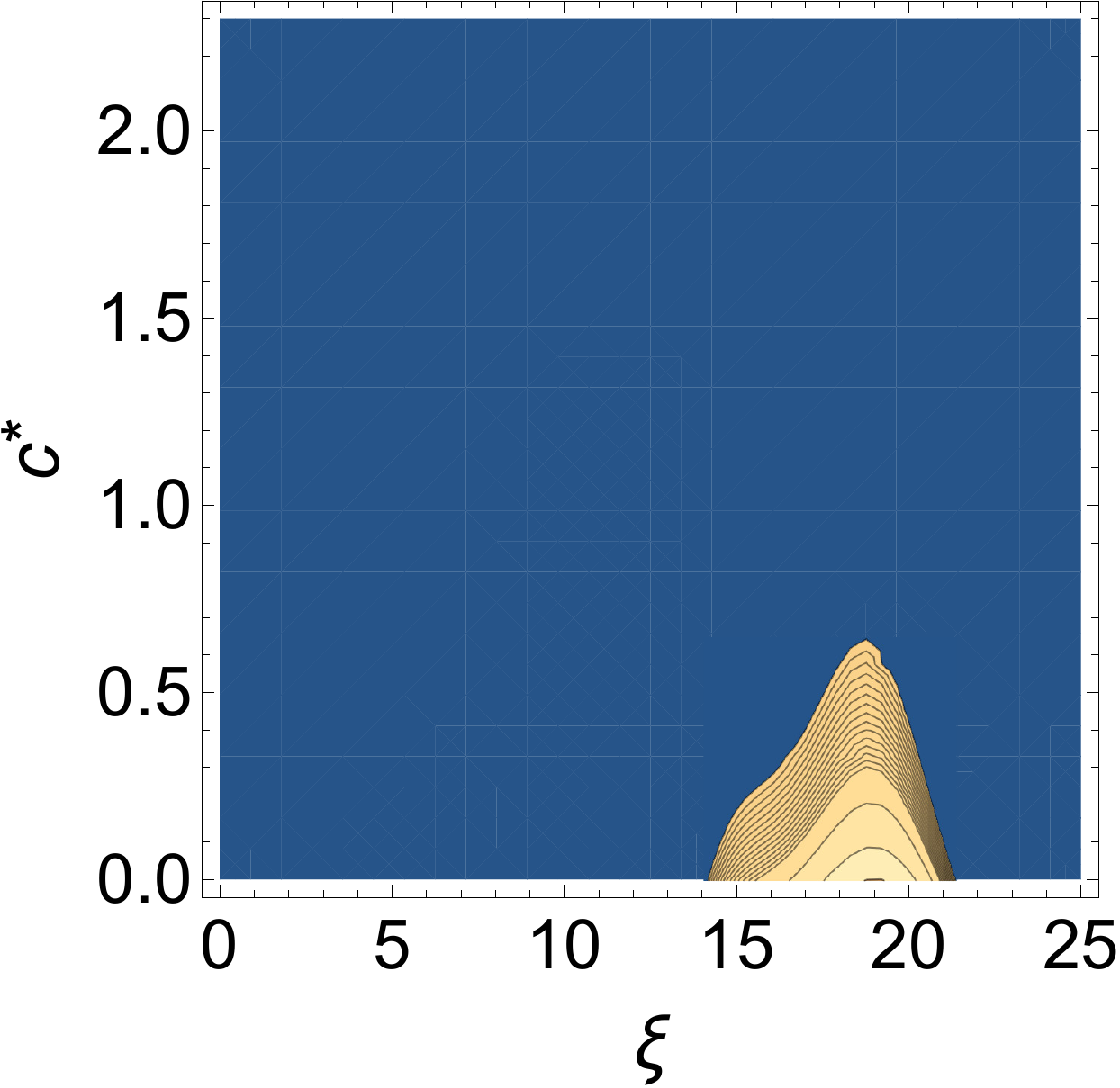}%
     }\\
     \subfloat[\label{subfig2g}]{%
       \includegraphics[width=.3\textwidth]{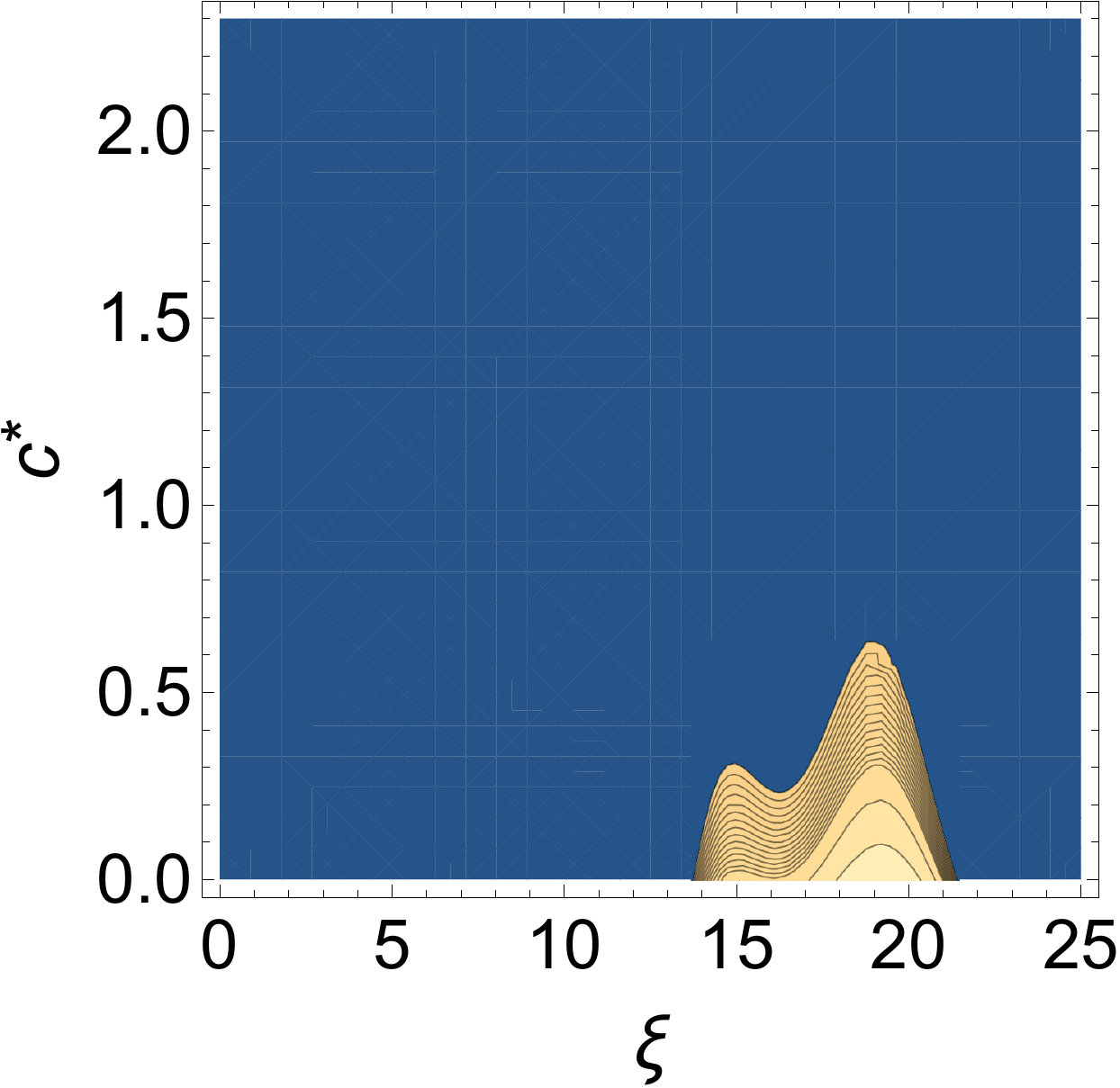}%{Lambda3ContourPlotqa022qr022}
     } \hfill
      \subfloat[\label{subfig2h}]{%
       \includegraphics[width=.3\textwidth]{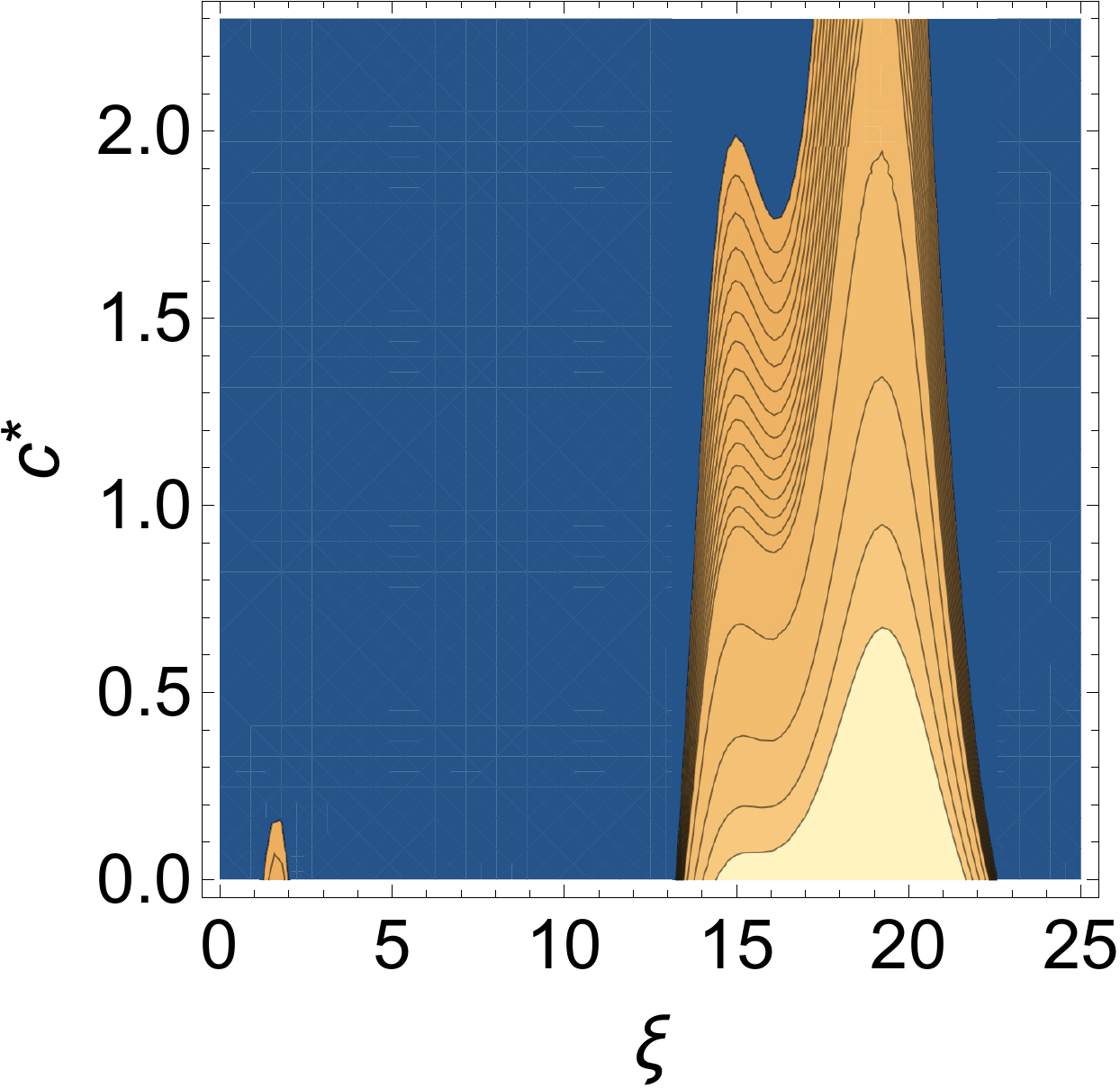}%
     }
     \hfill
     \subfloat[\label{subfig2i}]{%
       \includegraphics[width=.3\textwidth]{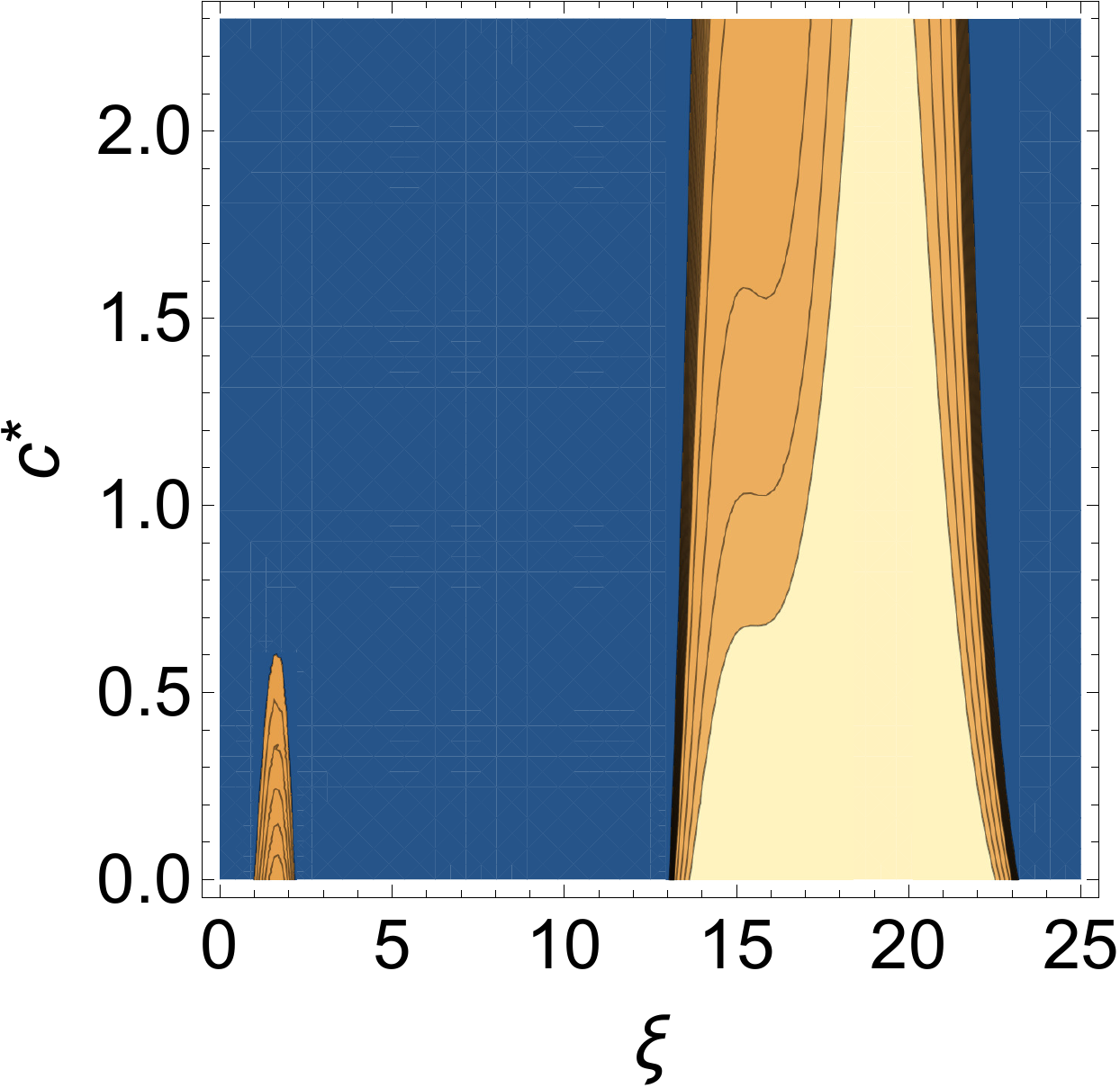}%{Lambda3ContourPlotqa022qr022}
     }
\caption{Contour plots of the dispersion relation $\lambda^1_u\left(\xi, c^{\star} \right)$ as $c^{\star}$ varies for: (a) $q_a=0.14$, $q_r=0.01$;  (b) $q_a=0.16$, $q_r=0.01$; (c) $q_a=0.22$, $q_r=0.01$; (d) $q_a=0.33$, $q_r=0.01$; (e) $q_a=0.01$, $q_r=0.22$; (f) $q_a=0.14$, $q_r=0.22$; (g) $q_a=0.22$, $q_r=0.22$; (h) $q_a=0.33$, $q_r=0.33$; (i) $q_a=0.44$, $q_r=0.44$. All other parameter values are given in Tables~\ref{tab:1} and ~\ref{tab:2}. Negative values of $\lambda_u^1(\xi,0)$ have been set to zero for better visualisation.}
\label{Fig_2}
\end{figure}

We next establish the effect of \ca on ADI. In Fig.~\ref{Fig_2} we display contour plots corresponding to non-negative values of $\lambda^1_u(\xi,c^{\star})$, for nine different combinations of $q_a$ and $q_r$:  (0.14, 0.01), (0.16, 0.01), (0.22, 0.01), (0.33, 0.01), (0.01, 0.22), (0.14, 0.22), (0.22, 0.22), (0.33, 0.33) and (0.44, 0.44).
We observe that the ADI regions vanish at sufficiently large values of $c^{\star}$ for all figures except Figs.~\subref*{subfig2d}, \subref*{subfig2h} and \subref*{subfig2i}. Note that we choose $0 \leq c^{\star}\leq 2.3$ since 2.3 is the maximum value of the steady state of the Atri model; other \ca models may achieve higher $c^{\star}$ levels but we expect a qualitatively similar behaviour. Also, as $c^{\star}$ increases the range of $\xi$ in the ADI regions decreases. In Figs.~\subref*{subfig2c},\subref*{subfig2d}, \subref*{subfig2h} and \subref*{subfig2i} we observe disjoint parameter regions for positive $\lambda^1_u(\xi,c^{\star})$.

%%%%%%%%%%%%%%%%%%%%%%%%%%%%%%%%%%%%%%---------- Fig 2 ----------%%%%%%%%%%%%%%%%%%%%%%%%%%%%%%%%%%%%%%

%%%%%%%%%%%%%%%%%%%%%%%%%%%%%%%%%%%%%%%%%%%%%%%%%%%%%%%%%%%%%%%%%%%%%%%%%%%%%%%%%%%%%%%%%%%%%%%%%%%%%%%%%%%%
The stability of the spatially homogeneous Atri model (determined by the matrix $J_2$) has been covered in detail in \citep{atri1993single, kaouri2019simple}. Hopf bifurcations occur at $\mu=0.289$ and $\mu=0.495$, between which stable relaxation oscillations (limit cycles) exist. Action potentials also appear for a very small range of $\mu$. Including diffusion leads to the emergence of periodic wave trains and solitary pulses when the Atri model exhibits limit cycles and action potentials, respectively (see Fig.~\ref{Fig_4}).

\section{Numerical simulations}
\label{numerics}

In this section we numerically solve model~\eqref{Nondim_AtriCellsDiff_nonlocal} using a method-of-lines approach.  The domain $[0, L]$ is discretized into a cell-centered grid with uniform length $h = 1/N$, where $N=100$ is the number of grid cells per unit length. All simulations are performed with $L=120$ and with periodic boundary conditions. The diffusion terms are discretized using a second order centered difference scheme. The advection term is discretized using a third order upwind scheme, augmented with a a flux-limiting scheme to ensure the solution's positivity.
The non-local term in equation~\eqref{Nondim_AtriCellsDiff_nonlocal;c} presents challenges regarding its efficient and accurate evaluation. Here we employ the scheme based on the Fast Fourier Transform introduced in \citep{gerisch2009approximation}, using the trapezoidal rule to pre-compute the integration weights. The resulting system of ODEs is integrated using the ROWMAP integrator introduced in \cite{weiner1996rowmap}.  We use the implementation\footnote{\url{http://www.mathematik.uni-halle.de/wissenschaftliches_rechnen/forschung/software/}} provided in \citep{weiner1996rowmap}. The integrator (written in Fortran) was wrapped using f2py into a scipy integrate class \citep{2019arXiv190710121V}. The spatial discretisation (right hand side of ODE) is implemented using NumPy. The integrator's error tolerance is set to $v_\mathrm{tol}=10^{-7}$. For the full details of the numerical methods we refer to \citep{gerisch2001numerical,Hundsdorfer}.

The initial conditions of the system are taken to be narrow Gaussian functions as follows:
\begin{subequations}
\label{ICs_numerics}
\begin{align}
& c\left(x,0\right)=c^{\star}+1.42857e^{-0.25\left(x-\frac{L}{2}\right)^2},\label{ICs_numerics;a}\\
& h\left(x,0\right)=\dfrac{1}{1+c^{{\star}^2}},\label{ICs_numerics;b}\\
& u\left(x,0\right)=e^{-0.25\left(x-\frac{L}{2}\right)^2}\label{ICs_numerics;c}.
\end{align}
\end{subequations}

\subsection{Adhesion-driven instability, pattern formation and cell aggregations}
Each term in the cancer cell density
equation~\eqref{Nondim_AtriCellsDiff_nonlocal;c} critically affects the
behaviour of cancer cells. Thus, below we examine the effect of each term in
turn and compare the results with those of the linear stability analysis in the
absence of \ca, in Section \ref{lin_stability}. We explore a wide range of
values for $q_a$ (magnitude of attraction) and $q_r$ (magnitude of repulsion),
guided by Fig. \ref{Fig_1}. For $q_a$ we also take into account the range of
$q_a$ reported in \eqref{admiss_values_qas}), based on measurements of the speed
of cancer cell movement. No experimental evidence was found for $q_r$ and we
consider the same range as for $q_a$. We thus examine several possible scenarios
and identify various types of patterns and aggregations.

\begin{figure}
    \centering
    \includegraphics[width=1\textwidth]{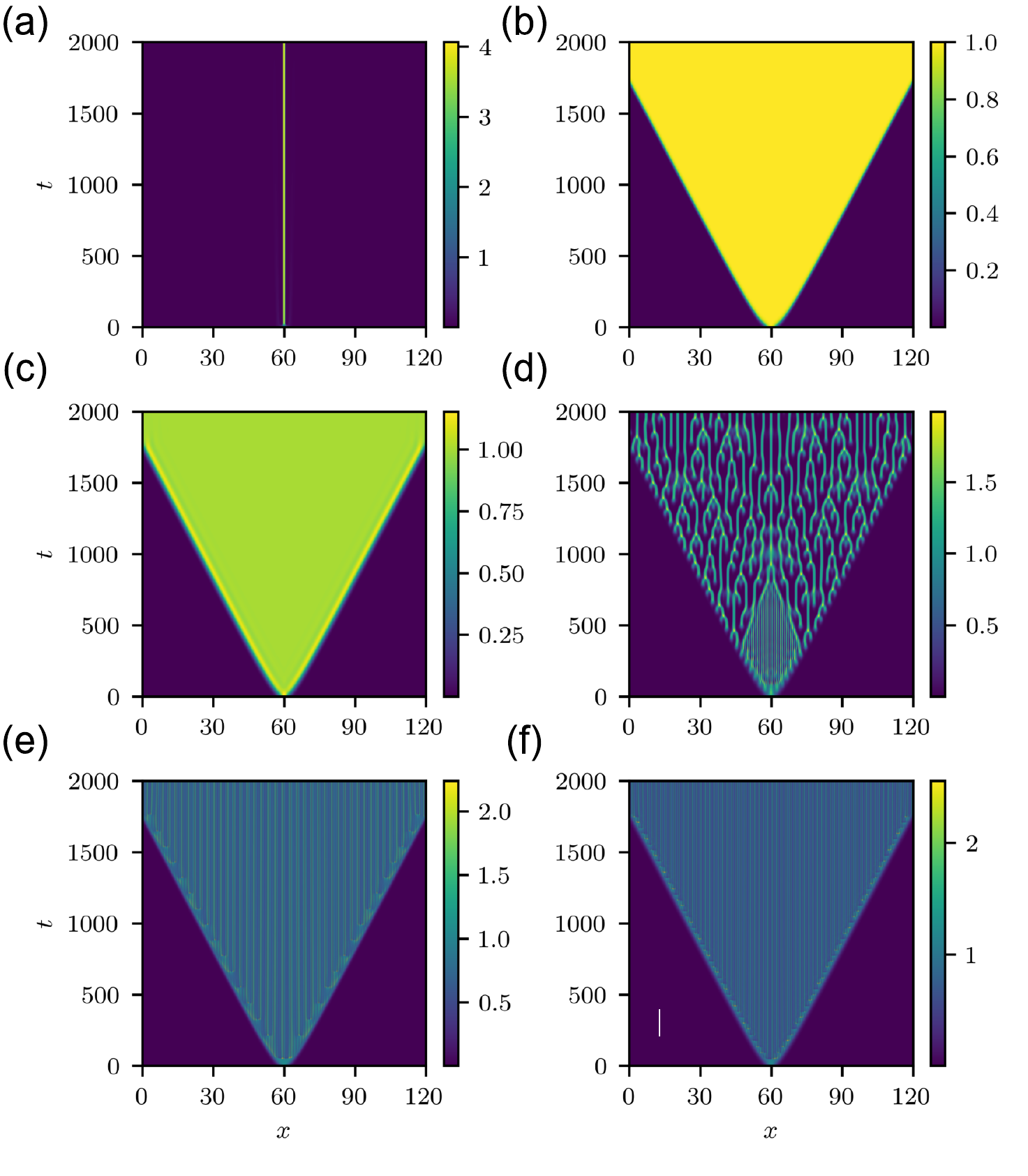}
    \caption{Cancer cell density, $u(x,t)$, for no \ca  effect ($a_1=a_2=r_2=r_3=0$), governed by equation~\eqref{Nondim_AtriCellsDiff_nonlocal;c}. The initial conditions are given in (\ref{ICs_numerics;c}). (a) $q_a=0.22, q_r=0.01$, no proliferation; (b) $q_a=0, q_r=0$;  (c) $q_a=0.14, q_r=0.01$; (d) $q_a=0.22, q_r=0.01$; (e) $q_a=0.14, q_r=0.22$ ; (f) $q_a=0.22, q_r=0.22$. All other parameter values as in  Tables~\ref{tab:1} and ~\ref{tab:2}.}
    \label{Fig_3}
\end{figure}

In Fig.~\ref{Fig_3}(a) we plot the cell density for non-zero diffusion and advection but zero proliferation; this represents cells with very slow doubling time. We take $q_a=0.22$, $q_r=0.01$, i.e. attraction much larger than repulsion. We see that the cancer cells form a single stationary pulse. In Fig.~\ref{Fig_3}(b), we add proliferation, but take zero adhesion (Fisher's equation). The cancer cells exhibit a travelling front that propagates in opposite directions at a constant speed, as expected \citep{murray2003ii}. In Fig.~\ref{Fig_3}(c) we include diffusion, advection and proliferation, with $q_a=0.14$ and $q_r=0.01$. We still see a Fisher-like travelling front, consistently with Fig.~\subref*{subfig1a} which predicts no ADI for these choices of $q_a$ and $q_r$.

In Fig.~\ref{Fig_3}(d), we further increase the strength of attraction to $q_a=0.22$ while keeping $q_r=0.01$, and a pattern emerges behind the travelling front due to ADI, as predicted by Fig.~\subref*{subfig1b}. It is a ``mixed"  pattern, featuring merging and emerging peaks; some cancer cells form stationary pulses, while others organise into travelling pulses. This behaviour can be explained by the strong attractive forces that make cells form large aggregations. This type of pattern has been identified in previous work (see \cite{andasari2011mathematical,bitsouni2017mathematical,hillen2009user,loy2019modelling,eftimie2017pattern,wang2007pattern}),
%In \cite{andasari2011mathematical} they have also investigated the effect of diffusion coefficient in this type of patterns, indicating the existence of this type of pattern for small diffusion or in other words for adhesion-driven instability.

%%%%%%%%%%%%%%%%%%%%%%%%%%%%%%%%%%%%%%---------- Fig 3 ----------%%%%%%%%%%%%%%%%%%%%%%%%
%%%%%%%%%%%%%%%%%%%%%%%%%%%%%%%%%%%%%%%%%%%%%%%%%%%%%%%%%%%%%%%%%%%%%%%%%%%%

% Let's try this
In Fig.~\ref{Fig_3}(e), we lower attraction to $q_a=0.14$ and increase the magnitude of repulsion to $q_r=0.22$; the Fisher-like front persists and the pattern behind it now exhibits thin spikes. This can be explained by the strong repulsive forces leading to a larger number of smaller aggregations than those in the case where attraction is larger than repulsion, as in Fig.~\ref{Fig_3}(d).This behaviour again agrees with the linear stability analysis (see Fig.~\ref{Fig_1}(b)). %This behaviour confirms the large critical wave number in Fig.~\ref{Fig_1}(a), which is larger that the critical wave number occurring in the case of large attraction in Fig.~\ref{Fig_1}(b).
Finally, in Fig.~\ref{Fig_3}(f) we take equal attraction and repulsion, $q_r=q_a=0.22$. The pattern is similar to that in Fig.~\ref{Fig_3}(e). Note: in order to see the more detailed features of the Figures the reader is encouraged to follow the electronic version of the paper.

\subsection{Calcium signals}
\begin{figure}[!htbp]
     \centering
\includegraphics[width=1\textwidth, height=0.75\textwidth]{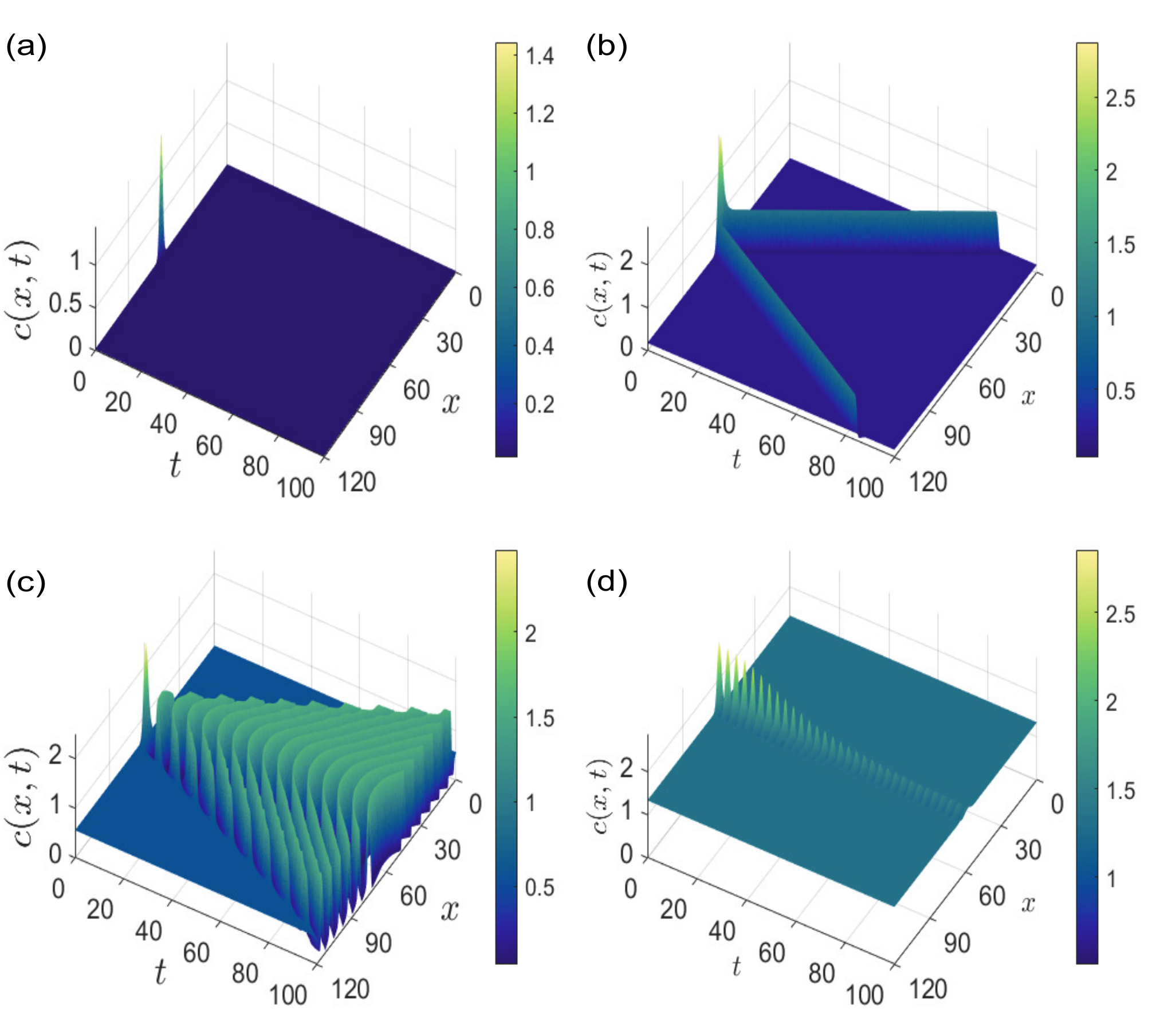}
    \caption{Patterns of the \ca concentration, $c(x,t)$, generated by the Atri model~\eqref{Nondim_AtriCellsDiff_nonlocal;a}-~\eqref{Nondim_AtriCellsDiff_nonlocal;b} for (a) $\mu=0.1$
    ($c^{\star}=0.016$), (b) $\mu=0.288$ ($c^{\star}=0.177$), (c) $\mu=0.3$ and (d) $\mu=0.5$ ($c^{\star}=1.332$). The initial conditions are given in (\ref{ICs_numerics;a})-(\ref{ICs_numerics;b}). The remaining  parameter values are given in  Tables~\ref{tab:1} and ~\ref{tab:2}. Note that although we report $c^{\star}$ for all $\mu$ when \ca is oscillatory the steady state is linearly unstable}
 \label{Fig_4}
   \end{figure}

Here, we investigate the behaviour of the spatially extended Atri model \eqref{Nondim_AtriCellsDiff_nonlocal;a} and \eqref{Nondim_AtriCellsDiff_nonlocal;b}. The four panels in Fig.~\ref{Fig_4} display the behaviour of the \ca concentration as we increase $\mu$, which is equivalent to increasing the \ip concentration. For $\mu=0.1$, for which the spatially clamped Atri model possesses a linearly stable fixed point \citep{atri1993single, kaouri2019simple}, Fig.~\ref{Fig_4}(a) illustrates that the initial Gaussian condition decays to this fixed point. Setting $\mu=0.288$ leads to a solitary travelling pulse (Fig.~\ref{Fig_4}(b)), while a value of $\mu$ between the two Hopf bifurcations results in a periodic wave train \citep{keener2009mathematical1,keener2009mathematical2}; in Fig.~\ref{Fig_4}(c) we take, as an example, $\mu=0.3$. Finally, for larger values of $\mu$ the Atri model is linearly stable again and we find a similar pattern to Fig.~\ref{Fig_4}(a), in that the initial condition decays to the steady state, but in a periodic manner. In Fig.~\ref{Fig_4}(d) we take $\mu=0.6$ as an example of the latter case. These four types of \ca signals emerge in almost all \ca models. Here, we use them as input to the cancer cell density equation \eqref{Nondim_AtriCellsDiff_nonlocal;c}.

%%%%%%%%%%%%%%%%%%%%%%%%%%%%%%%%%%%%%%---------- Fig 4 ----------%%%%%%%%%%%%%%%%%%%%%%%%%%%%%%%%%%%%%%%

%%%%%%%%%%%%%%%%%%%%%%%%%%%%%%%%%%%%%%
%\todo[inline]{Andreas can you run for calcium only in proliferation ($a_1=a_2=0$) and then calcium only in adhesion ($r_1=r_2=0$)?}
%\todo[inline]{Amplitude to be discussed}

\subsection{The effect of \ca on the cell density}
\begin{figure*}[!htbp]
 \centering
\includegraphics[width=1\textwidth]{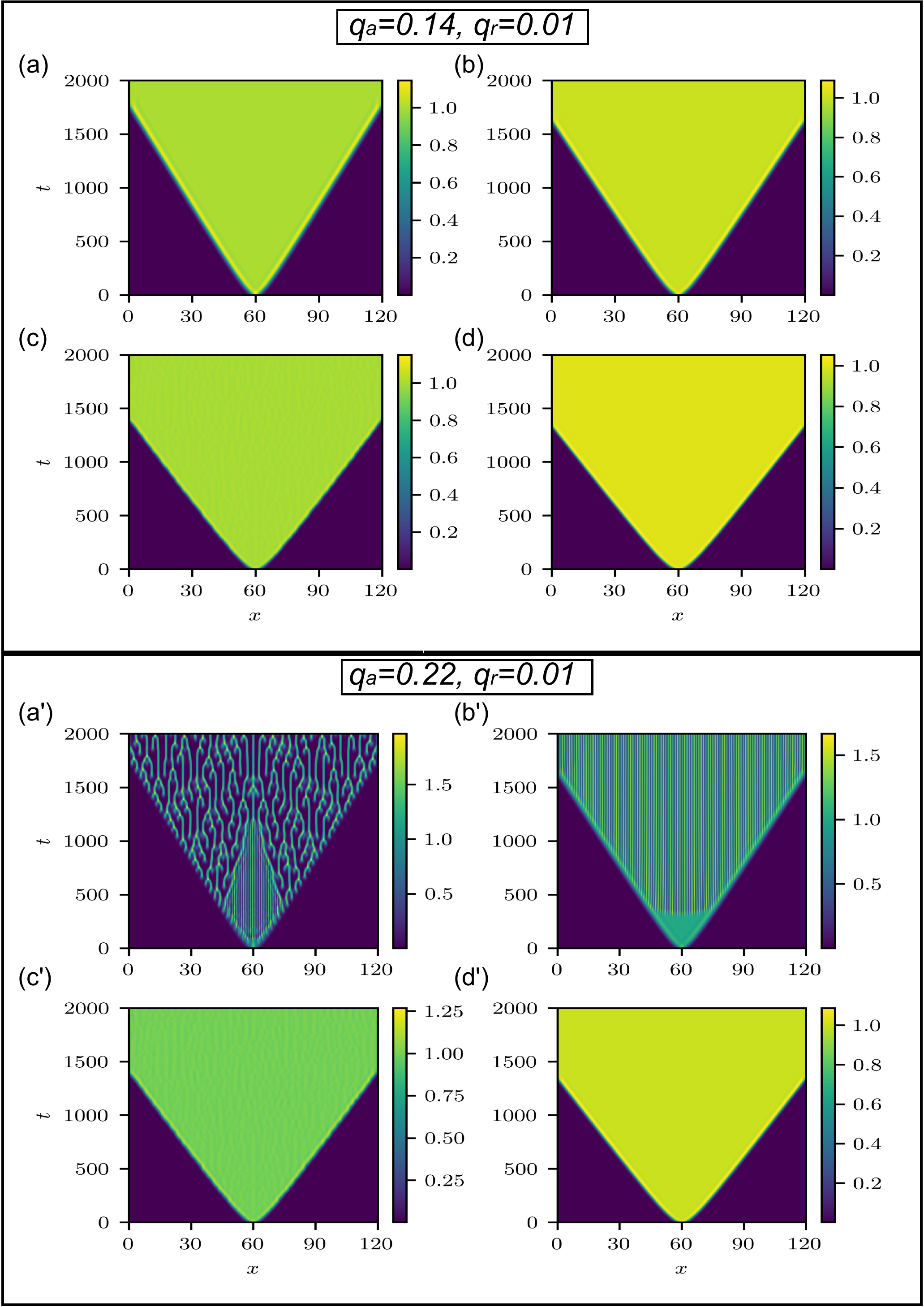}
\caption{Cancer cell density, $u(x,t)$, governed by
equation~\eqref{Nondim_AtriCellsDiff_nonlocal;c}, as  $q_a$ increases (top panel
for $q_a=0.14$ and bottom panel for $q_a=0.22$); $q_r=0.01$. The initial
conditions are given by (\ref{ICs_numerics}). For (a), (a$^\prime$) $\mu=0.1$,
$c^{\star}=0.016$ (non-oscillatory \ca); (b), (b$^\prime$)  $\mu=0.3$
($c^{\star}=0.556$, oscillatory \ca); (c), (c$^\prime$)  $\mu=0.45$
($c^{\star}=1.195$, oscillatory \ca); (d), (d$^\prime$) $\mu=0.6$
$c^{\star}=1.5712$  (non-oscillatory \ca). The rest of model parameters are
given in  Tables~\ref{tab:1} and~\ref{tab:2}. As predicted from the linear
stability analysis (see Fig.~\protect\subref*{subfig2b}), when $q_a$ increases ADI
emerges for small values of $\mu$. Note that although we report $c^{\star}$ for
all $\mu$ when \ca is oscillatory the steady state is linearly unstable.}\label{Fig_5}
\end{figure*}

We now examine the effect of the \ca signals on the cancer cell density.  We fix the attraction and repulsion magnitudes, $q_a$ and $q_r$, and vary $\mu$.
Fig.~\ref{Fig_5} (top panel) ($q_a=0.14$, $q_r=0.01$) shows a Fisher-like travelling front in all Figs.(a)--(d), irrespective of the \ip and \ca  levels; this is consistent with the linear stability analysis that predicts no ADI. These results are in line with Fig.~\subref*{subfig2a}. In contrast, when we increase $q_a$ to $0.22$ in Fig.~\ref{Fig_5} (bottom panel) small \ip concentrations ($\mu=0.1$ and $\mu=0.3$, respectively) induce a pattern, due to ADI. As we increase the \ip concentration, the pattern vanishes, as illustrated in Figs.~\ref{Fig_5}(c$^\prime$) and \ref{Fig_5}(d$^\prime$) which are for $\mu=0.45$ and $\mu=0.6$, respectively. These results are in line with Fig.~\subref*{subfig2c}.
%%%%%%%%%%%%%%%%%%%%%%%%%%%%%%%%%%%%%%---------- Fig 5 ----------%%%%%%%%%%%%%%%%%%%%%%%%%%%%%%%%%%%%%%
%%%%%%%%%%%%%%%%%%%%%%%%%%%%%%%%%%%%%%%%%%%%%%%%%%%%%%%%%%%%%%%%%%%%%%%%%%%%%%%%%%%%%%%%%%%%%%%%%%%%%%%

%\subsection{The effect of \ca on ADI}

%\todo[inline]{VB: Maybe we should call them merging and emerging patterns/ local peaks (also called coarsening process), instead of `tree-type' patterns}

\begin{figure*}[!htbp]
 \centering
\includegraphics[width=1\textwidth]{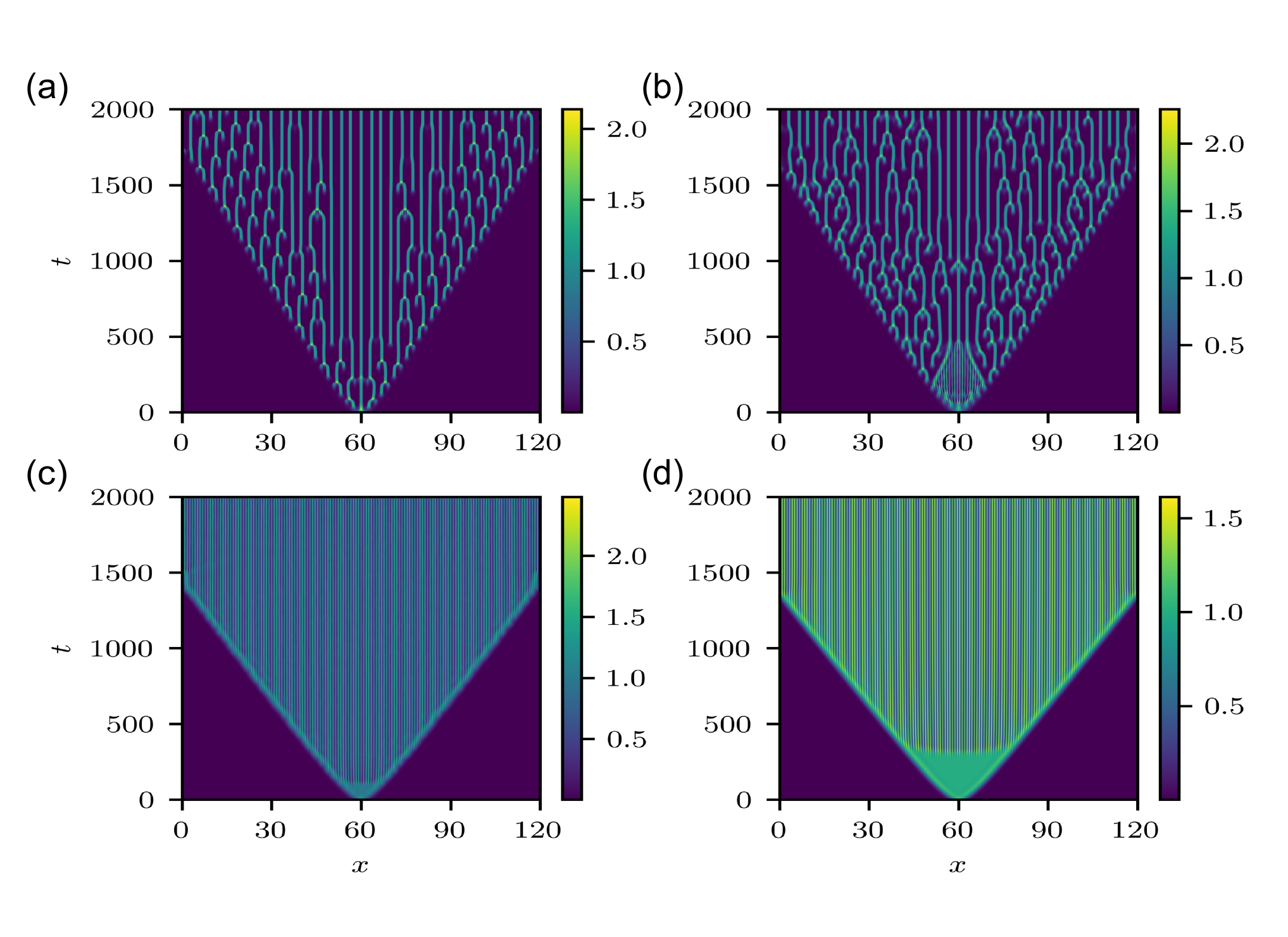}
\caption{Cancer cell density, $u(x,t)$, governed by equation~\eqref{Nondim_AtriCellsDiff_nonlocal;c}, for $q_a=0.33$, $q_r=0.01$. The initial conditions are given in (\ref{ICs_numerics}). (a) $\mu=0.1$, $c^{\star}=0.016$ (non-oscillatory \ca); (b) $\mu=0.3$, $c^{\star}=0.556$  (oscillatory \ca); (c) $\mu=0.45$, $c^{\star}=1.195$  (oscillatory \ca); (d) $\mu=0.6$, $c^{\star}=1.5712$  (non-oscillatory \ca). The rest of model parameters are given in  Tables~\ref{tab:1} and ~\ref{tab:2}. Note that although we report $c^{\star}$ for all $\mu$ when \ca is oscillatory the steady state is linearly unstable.}
\label{Fig_6}
\end{figure*}

In Figs.~\ref{Fig_6} we see that for larger values of $q_a$ ($q_a=0.33$, $q_r=0.01$) patterns emerge behind the Fisher-like front for all values of $\mu$. This is consistent with the linear stability analysis --- see Fig.~\subref*{subfig2d}. For small values of $\mu$, $\mu=0.1$ and $\mu=0.3$ in Figs.~\ref{Fig_6}(a) and ~\ref{Fig_6}(b), respectively,  the cancer cells exhibit merging and emerging peaks; cells move towards each other forming new aggregations of new cells and of cells that broke off from existing aggregations and in the long-term dynamics stationary pulses are also formed. \citep{bitsouni2017mathematical,loy2019modelling,eftimie2017pattern,wang2007pattern}. For larger values of $\mu$, and consequently larger values of \ca (see Figs.~\ref{Fig_6}(c) and ~\ref{Fig_6}(d)) the patterns are thin stripes (stationary pulses).

%%%%%%%%%%%%%%%%%%%%%%%%%%%%%%%%%%%%%%---------- Fig 6 ----------%%%%%%%%%%%%%%%%%%%%%%%%%%%%%%%%%%%%%%
%%%%%%%%%%%%%%%%%%%%%%%%%%%%%%%%%%%%%%%%%%%%%%%%%%%%%%%%%%%%%%%%%%%%%%%%%%%%%%%%%%%%%%%%%%%%%%%%%%%%%%%

\begin{figure*}[!htbp]
 \centering
\includegraphics[width=1\textwidth]{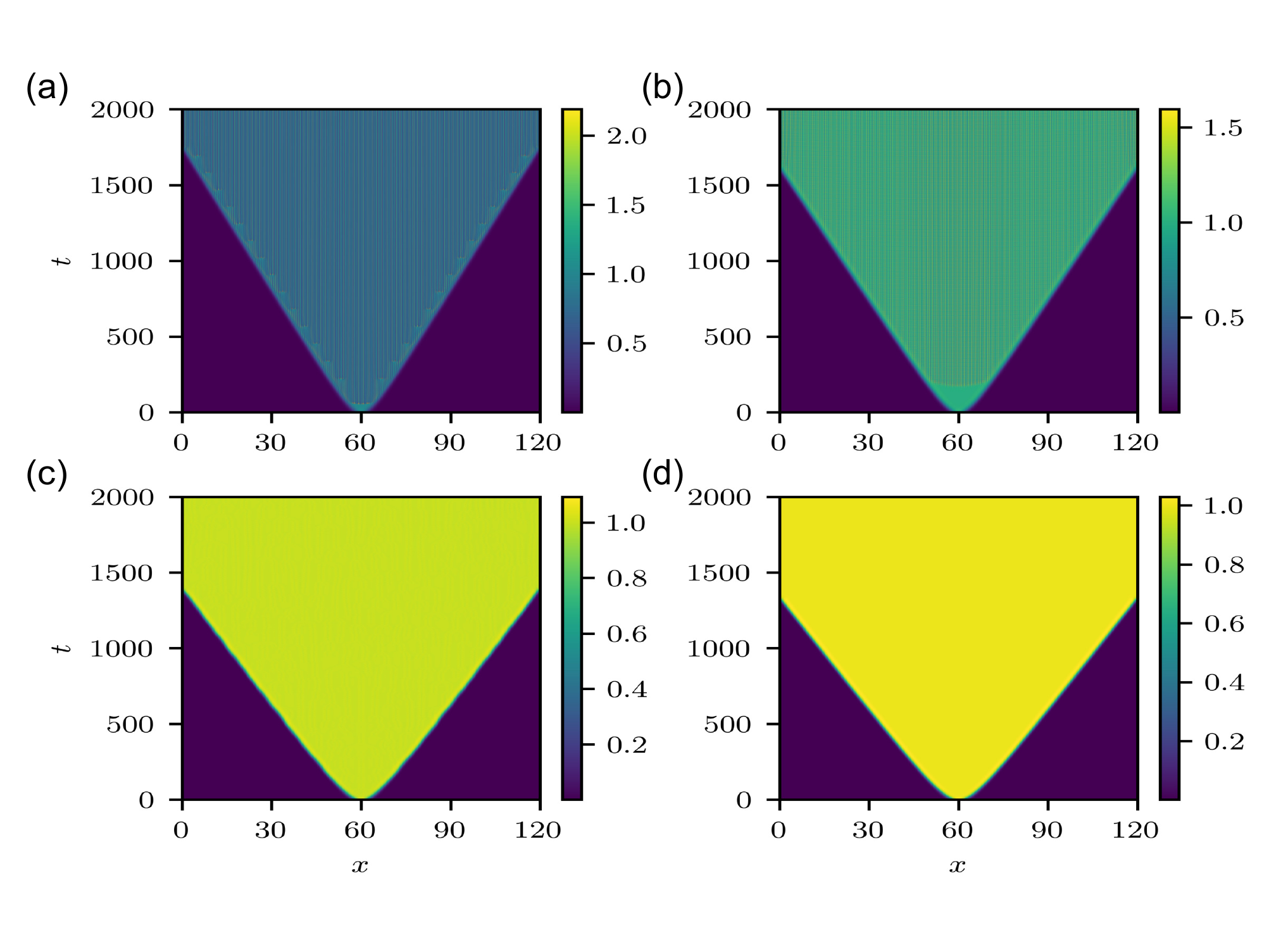}
\caption{Cancer cell density, $u(x,t)$, governed by equation~\eqref{Nondim_AtriCellsDiff_nonlocal;c}, for $q_a=0.14$ and $q_r=0.22$. The initial conditions are given in (\ref{ICs_numerics}). (a) $\mu=0.1$, $c^{\star}=0.016$ (non-oscillatory \ca); (b) $\mu=0.3$, $c^{\star}=0.556$  (oscillatory \ca); (c) $\mu=0.45$, $c^{\star}=1.195$  (oscillatory \ca); (d) $\mu=0.6$, $c^{\star}=1.5712$  (non-oscillatory \ca). The rest of model parameters are given in  Tables~\ref{tab:1} and ~\ref{tab:2}. Note that although we report $c^{\star}$ for all $\mu$ when \ca is oscillatory the steady state is linearly unstable.}
\label{Fig_7}
\end{figure*}

\begin{figure*}[!htbp]
 \centering
\includegraphics[width=1\textwidth]{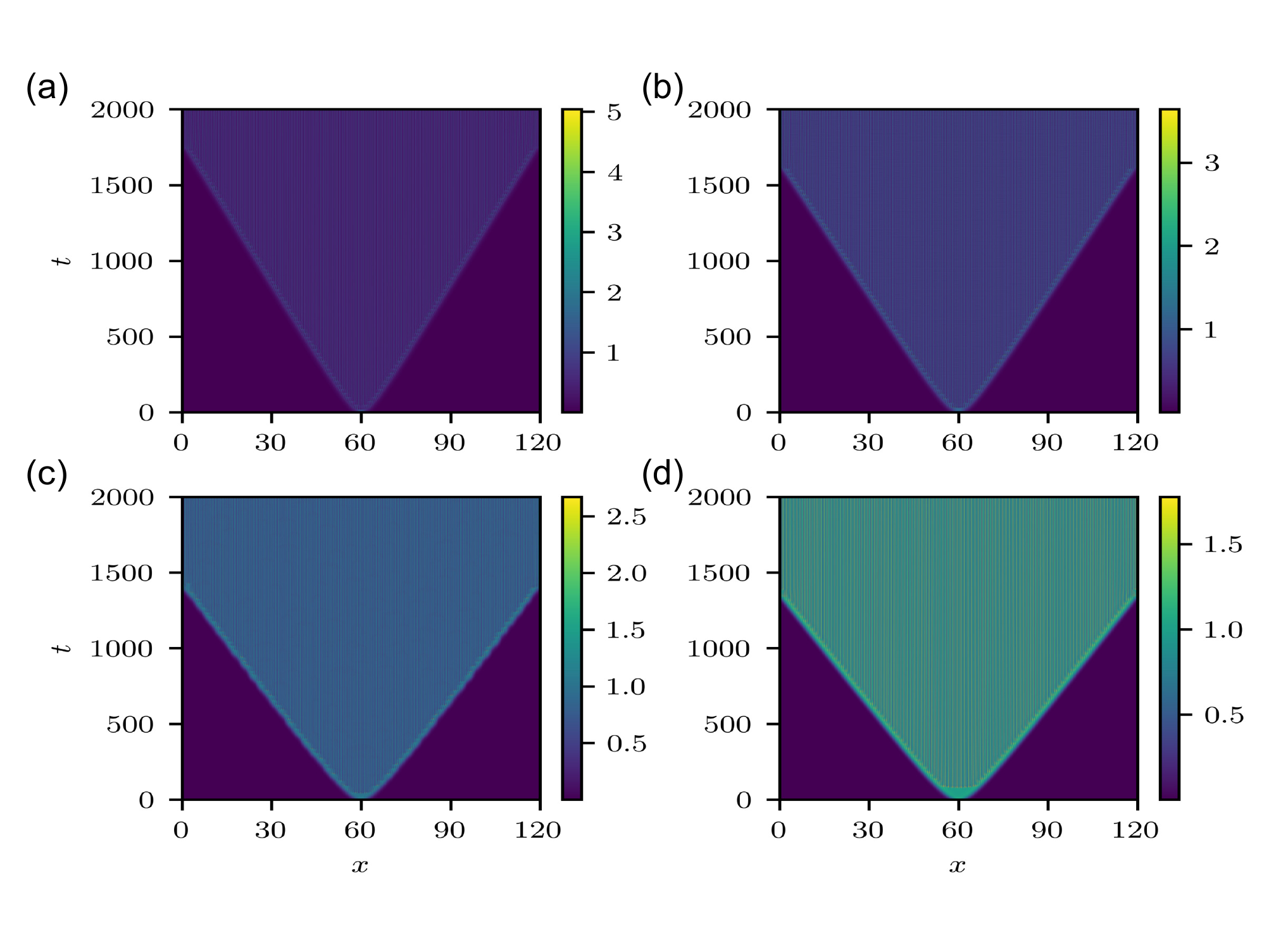}
\caption{Cancer cell density, $u(x,t)$, governed by
equation~\eqref{Nondim_AtriCellsDiff_nonlocal;c},for $q_a=q_r=0.33$. The initial
conditions are given in (\ref{ICs_numerics}). (a) $\mu=0.1$, $c^{\star}=0.016$
(non-oscillatory \ca); (b) $\mu=0.3$, $c^{\star}=0.556$  (oscillatory \ca); (c)
$\mu=0.45$, $c^{\star}=1.195$  (oscillatory \ca); (d) $\mu=0.6$,
$c^{\star}=1.5712$  (non-oscillatory \ca). The rest of model parameters are
given in  Tables~\ref{tab:1} and ~\ref{tab:2}. The results are consistent with
Fig.~\protect\subref*{subfig2g}. Note that although we report $c^{\star}$ for all $\mu$
when \ca is oscillatory the steady state is linearly unstable.} \label{Fig_8}
\end{figure*}

In Figs.~\ref{Fig_5} and \ref{Fig_6} attraction dominates over repulsion. In Fig.~\ref{Fig_7} we plot the cancer cell density when repulsion is stronger than attraction ($q_a=0.14$, $q_r=0.22$). For small values of the \ip concentration ($\mu=0.1$, $\mu=0.3$), Figs.~ \ref{Fig_7}(a),(b) exhibit thin-stripe patterns via ADI (stationary pulses). As we increase $\mu$, patterns vanish --- see Figs.~\ref{Fig_7}(c) and \ref{Fig_7}(d), respectively for $\mu=0.45$ and $\mu=0.6$. These results are consistent with Fig.~\subref*{subfig2f}.
%%%%%%%%%%%%%%%%%%%%%%%%%%%%%%%%%%%%%%---------- Fig 7 ----------%%%%%%%%%%%%%%%%%%%%%%%%%%%%%%%%%%%%%%
%%%%%%%%%%%%%%%%%%%%%%%%%%%%%%%%%%%%%%%%%%%%%%%%%%%%%%%%%%%%%%%%%%%%%%%%%%%%%%%%%%%%%%%%%%%%%%%%%%%%%%%
Finally, for large and equal values of $q_a$ and $q_r$, Figs.~\ref{Fig_2}(g) and ~\ref{Fig_2}(h) predict that ADI patterns exist for all \ca concentrations within the physiological range of the Atri model. This is confirmed in Fig.~\ref{Fig_8}, where we observe ADI patterns for any \ip (and \ca) level when $q_a=q_r=0.33$.
%%%%%%%%%%%%%%%%%%%%%%%%%%%%%%%%%%%%%%---------- Fig 8 ----------%%%%%%%%%%%%%%%%%%%%%%%%%%%%%%%%%%%%%%

Above, we have established the emergence and disappearance of patterns as \ca
varies. Furthermore, below we will summarise the effect of \ca on three
important characteristics of the solution: the wave speed of the Fisher-like
front, and also the amplitude and frequency of the cancer cell density.
%%%%%%%%%%%%%%%%%%%%%%%%%%%%%%%%%%%%%%%%%%%%%%%%%%%%%%%%%%%%%%%%%%%%%%%%%%%%%%%%%%%%%%%%%%%%%%%%%%%%%%%

\paragraph{Wave speed:}
In Figs.~\ref{Fig_5}--\ref{Fig_8} we see that as $\mu$ increases (fixed $q_a$ and $q_r$) the speed of the travelling front increases. This can be linked to a higher invasion and hence metastatic potential of the cancer cells. On the other hand, for fixed $\mu$ the wave speed does not change much as $q_a$ and/or $q_r$ vary. %IT WOULD BE GOOD TO HAVE SOME QUANTITATIVE ESTIMATES HERE-KK TO DO IF THERE IS TIME.

\paragraph{Amplitude:} Comparing Figs.~\ref{Fig_5} and \ref{Fig_6} we see that the maximal cell density increases as $q_a$, the attraction magnitude, increases from $0.14$ to $0.33$. Also, comparing Figs. \ref{Fig_6} and \ref{Fig_8}  we see a significant increase in the maximal cell density as $q_r$ increases from $0.01$ to $0.33$ (and $q_a$ fixed to $0.33$). The same effect is observed when comparing Fig. \ref{Fig_5}-top panel with Fig.\ref{Fig_7}, where again $q_r$ increases from $0.01$ to $0.33$ (while $q_a$ is fixed to $0.14$.). %Examining the cell density increase in more detail, we took only the adhesion to be \ca-dependent and then only the proliferation so as to decouple these two \ca-dependent effects. We found that in both decoupled cases the cell density increases as $q_a$ increases so adhesion and proliferation act synergistically in this.
For fixed $q_a$ and $q_r$ as $\mu$ increases the maximal cell density decreases, as we can see in Figs.~\ref{Fig_5}--\ref{Fig_8}.

\begin{figure*}[!htbp]
\centering
\includegraphics[width=1.\textwidth]{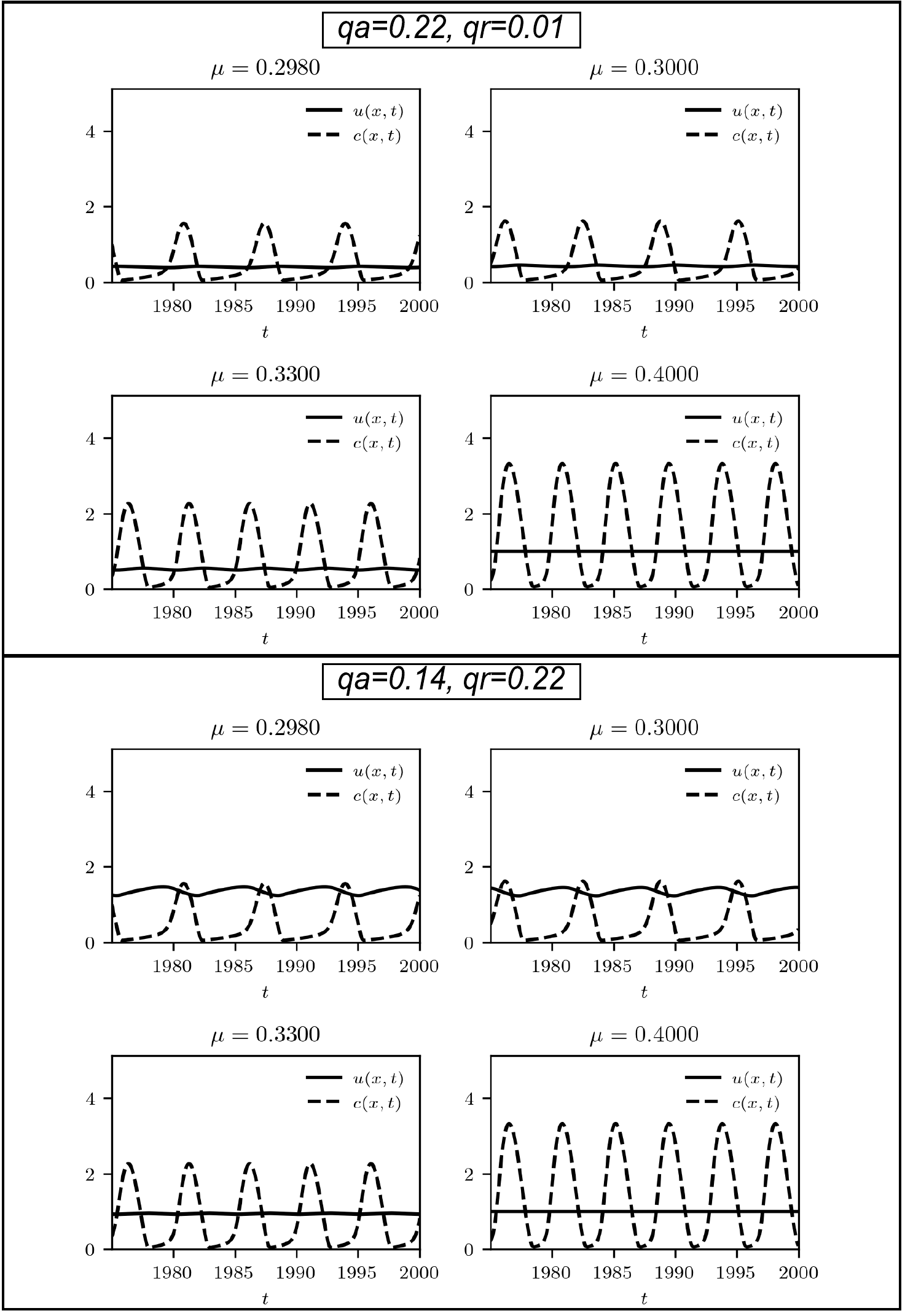}
\caption{Cancer cell density and \ca oscillations. Each plot shows a cross-section (i.e.\ $u(t) = u(55, t)$) of a solution of model~\eqref{Nondim_AtriCellsDiff_nonlocal} with initial conditions given in (\ref{ICs_numerics}) for selected increasing values of $\mu$. (Top)
$q_a=0.22$ and $q_r=0.01$. (Bottom) $q_a=0.22$ and $q_r=0.01$. The rest of model parameters are given in  Tables~\ref{tab:1} and ~\ref{tab:2}. The cell density $u(55, t)$ picks up the oscillations in the \ca concentration. Indeed the frequencies (computed using the Fourier transform) match.}
\label{Fig_9}
\end{figure*}

\paragraph{Frequency:}
Moreover, we investigate how \ca signalling affects the temporal frequency of cancer cell density oscillations. In Fig.~\ref{Fig_9} we fix $x=55$ and plot $c(x,t)$ and $u(x,t)$ for two choices; at the top panel we have $q_a=0.22$, $q_r=0.01$ (attraction much larger than repulsion) and in the bottom panel we have  $q_a=0.14$, $q_r=0.22$ (attraction comparable to repulsion). From the frequency bifurcation diagram of the Atri model (see Fig. 2 in \cite{kaouri2019simple}) we choose four values of $\mu$ that sufficiently `sample' the variation of the frequency as $\mu$ increases. We see that the frequency of \ca oscillations  is approximately equal to the frequency of cell density oscillations, \textbf{if} the cell density is oscillatory. We have verified this observation by also computing the frequency spectra for $t \in (1900, 2000)$ (the time interval has been chosen to ensure that solutions converged to steady state). For other choices of $q_a$ and $q_r$, the effect of \ca oscillations on the cell density is similar, and thus other figures are not included for brevity. %fix qa, qr in top panel change mu from 0.3 to 0.33 we see phase shit we do not see this for bottom panel. we did not observe phase shit in other choices of qa and qr

%\textcolor{red}{\the\textheight}
%\textcolor{red}{\the\textwidth}

%%%%%%%%%%%%%%%%%%%%%%%%%%%%%%%%%%%%%%---------- Fig 9 ----------%%%%%%%%%%%%%%%%%%%%%%%%%%%%%%%%%%%%%%

\section{Summary, conclusions and further work}
\label{conclusion}
Since cell proliferation and cell-cell adhesion, which play a critical role in invasion and cancer metastasis, are \ca-dependent, here we have developed and analysed a new model for \ca signalling in cancer. The \ca dynamics have been described by the spatially extended Atri model \citep{atri1993single}, which consists of a reaction-diffusion equation for the \ca concentration, coupled with an ODE for the fraction of \ip receptors on the ER that have not been inactivated by \ca. This model, although simple enough, generates four ‘prototypical’ \ca signals as many other excitable \ca models; periodic wavetrains (which correspond to limit cycles in the spatially clamped Atri model), solitary pulses (which correspond to action potentials), decaying wavetrains and solutions decreasing monotonically with time.  The cancer cell density evolution is described by a non-local PDE that incorporates diffusion, cell-cell adhesion (advection) and proliferation. We have modelled the dependence of the adhesion and proliferation terms on the \ca dynamics, motivated by experimental evidence, and we have considered cancer types where the adhesion strength decreases with \ca \citep{byers1995role,cavallaro2004cell}, while proliferation increases with \ca \citep{cardenas2016selective,prevarskaya2018ion,rezuchova2019type,tsunoda2005inositol}. The model, assumptions and parameter values are presented in Section~\ref{model_derivation}. As much as possible, the model parameters were chosen from experimental studies (see Tables~\ref{tab:1} and ~\ref{tab:2}).

In Section~\ref{analytical_results} we linearised the model \eqref{Nondim_AtriCellsDiff_nonlocal} and determined the parameter range for which an adhesion-driven instability (ADI) forms, while varying systematically the magnitudes of cell-cell attraction and repulsion, $q_{a}$ and $q_{r}$, respectively. In the absence of \ca (Fig.~\ref{Fig_1}) we showed that ADIs may arise for sufficiently large values of either $q_a$ and $q_r$ (or both). ADIs correspond to cell aggregations which are critical for cancer invasion and metastasis. Then, in Fig.~\ref{Fig_2} we investigated the effect of \ca on the cell aggregations and found that they change qualitatively and eventually vanish as the \ca level increases.

In Section~\ref{numerics} we solved the full non-linear model \eqref{Nondim_AtriCellsDiff_nonlocal} numerically and systematically investigated a range of attraction and repulsion magnitudes, guided by the linear stability analysis. Firstly, we validated numerically the results of the linear analysisin the absence of \ca (Fig.~\ref{Fig_3}). We subsequently examined the effect of four types of \ca signals on the cancer cell density, paying special attention to the periodic wave trains (Figs.~\ref{Fig_5}-\ref{Fig_8}). We found that as \ca levels increase the maximal cell density  decreases due to the decreased cell-cell adhesion strength preventing the formation of clusters of high density levels. %To confirm this result we reran simulations for explicitly \ca-dependent adhesion (ignoring the effect of \ca in proliferation) and showed adhesion-dominated movement and aggregation of cancer cells.
Moreover, as \ca levels increase the speed of the travelling wave fronts increases which is linked to a faster spread of cancer. An other important result from our numerical investigations is that the frequency of \ca oscillations is approximately equal to the frequency of the cancer cell density oscillations, when the cell density is oscillatory. Moreover, cellular aggregations vanish for sufficiently large  \ca levels, as it was predicted by the linear analysis. Our results demonstrate that accounting for the dependence of cell-cell adhesion and proliferation on \ca signalling we can reveal the conditions for which cancer cell aggregations appear as \ca varies. This allows us to study the dependence of the cancer invasion potential on \ca and paves the ways for new therapies based on controlling \ca.
%KK: @RT -should we report the wave speed values and temporal frequencies?

Our model provides a general framework for cancer cell movement under the effect of any oscillatory signalling pathway dynamics and paves the way for treatments that are based on controlling these pathways, and in particular \ca signalling. It, however, has various limitations which outline avenues for future work. The assumption that the adhesion strength function is decreasing with \ca is not appropriate for all cancer types; an increase of cell-cell adhesion with \ca has been observed in some cancers. Additionally, the repulsion magnitude has been taken over a wide range since there is no experimental evidence supporting its value. New experiments could investigate this. Another limitation of the model is that it includes cell–cell interactions; it would be useful to incorporate the interaction of the cancer cells with the extracellular matrix (ECM) in future work as this would allow to study cancer invasion in more detail. Additionally, the way cell-ECM interactions are dependent on \ca could be also modelled. Finally, the delay of the \ca waves in the gap junctions between cells has been considered negligible; a cell-based model accounting for these gap junctions could be developed. Moreover, as we are now equipped with the insights generated by the one-dimensional geometry, we plan to develop the model to two and three dimensions.

A main focus of this study was to unravel the impact of the cellular \ca signalling on the behaviour of cancer cells. As such, a key component of our model is the description of the cellular \ca dynamics. We chose the Atri model as a typical representative for a minimal framework that captures essential features of the dynamics of the cellular \ca concentration such as \ca oscillations. This naturally raises the question about how robust our results are with respect to the \ca model that we employed. The answer to this question combines two main lines of argument: the specific model for the \ipr and whether \ca oscillations are deterministic or stochastic. For the first point, we note that there exist a substantial number of \ipr models, see e.g. \citep{atri1993single,de1992single,li1994calcium,li1994equations,meyer1988molecular,Sneyd:2002fu,Siekmann:2012jl,Sneyd:2005ffa,Shuai:2007bk, Ullah:2012jr}. While they differ in their complexity, the overall range of the \ca concentration and the frequency of the \ca oscillations are comparable amongst them. Consequentially, exchanging the Atri model for any of the other \ipr models will most probably not change our conclusions. A more contentious point is whether \ca oscillations should be described within a deterministic or stochastic framework. Both approaches have been used extensively to date as e.g. in \citep{Dupont:2011ky, Falcke:2018jr, Gaspers:2014hc, Kummer:2000jg, Li:1994dh,Politi:2006en, Powell:2020ji, Shuai:2002ks,Skupin:2008hg, Sneyd:2017fs, Sun:2017gu, Tang:1996dy, Thul:2009db,Thul:2007gy,Thul:2006wv,Thul:2004es,Thul:2004tt, Thurley:2011jc,Tilunaite:2017kla,Thul:2014hy,Thurley:2012hf, TsanevaAtanasova:2005km, Voorsluijs:2019bu, Weinberg:2014hd, Wieder:2015fe} --- see also the book by \citep{dupont2016models} for a detailed discussion. As this study is the first to explore the role of \ca in a mathematical model of cancer cell propagation, we opted for a deterministic approach. This provides us with a baseline against which we can test future models in which the \ca dynamics will be described stochastically.

%%%%%%%%%%%%%%%%%%%%%%%%%%%%%%%%%%%%%%%%%%%%%%%%%%%%%%%%%%%%%%%%%%%%%%%%%%%%%%%%%%%%%%%%%%%%%%%%%%%%%%%

\begin{acknowledgements}

The authors would like to thank Dr. A. Athenodorou for his valuable technical support. VB acknowledges support from the European Union’s H2020 Research and Innovation Action under Grant Agreement No 741657 \href{https://www.scishops.eu/}{{\color{blue!30!black} (SciShops.eu)}}.
AB was partially supported by an NSERC (Natural Sciences and Engineering Research Council) post-doctoral fellowship, and is grateful to the Pacific Institute for Mathematical Sciences for providing space and resources for AB’s postdoctoral research.

\end{acknowledgements}

%%%%%%%%%%%%%%%%%%%%%%%%%%%%%%%%%%%%%%%%%%%%%%%%%%%%%%%%%%%%%%%%%%%%%%%%%%%%%%%%%%%%%%%%%%%%%%%%%%%%%%%

\bibliographystyle{spbasic}      % basic style, author-year citations
\bibliography{Biblio_Calcium_cancer}  % name your BibTeX data base

\end{document}